\documentclass[prb,a4paper,showpacs,twocolumn,superscriptaddress,longbibliography]{revtex4-1}

\usepackage{amssymb}
\usepackage{amsmath}
\usepackage{amsfonts}
\usepackage{graphicx}
\usepackage{bm}
\usepackage{color}
\usepackage{multirow}
\usepackage{natbib}
\usepackage{hyperref}

\renewcommand{\Re}{\,\textrm{Re}\,}
\renewcommand{\Im}{\,\textrm{Im}\,}

\DeclareMathOperator{\tr}{tr}

\begin{document}

\title{Charge relaxation resistance in the cotunneling regime of multi-channel Coulomb blockade: Violation of Korringa-Shiba relation}

\author{I. S. Burmistrov}

\affiliation{L.D. Landau Institute for Theoretical Physics RAS,
Kosygina street 2, 119334 Moscow, Russia}
\affiliation{Moscow Institute of Physics and Technology, 141700 Moscow, Russia}

\author{Ya. I. Rodionov}

\affiliation{Institute for Theoretical and Applied Electrodynamics RAS, Izhorskaya Str. 13, 125412 Moscow, Russia}

\begin{abstract}
We study the low frequency admittance of a small metallic island coupled to a gate electrode and to
a massive reservoir via a \emph{multi channel} tunnel junction. The ac current is caused by a slowly oscillating gate voltage. We focus on the regime of inelastic cotunneling in which the dissipation of energy (the real part of the admittance) is determined by two-electron tunneling with creation of electron-hole pairs on the island. We demonstrate that at finite temperatures but low frequencies the energy dissipation is ohmic whereas at zero temperature it is super-ohmic. We find that (i) the charge relaxation resistance (extracted from the real part of the admittance) is strongly temperature dependent, (ii) the imaginary and real parts of the admittance do not satisfy the Korringa-Shiba relation. At zero temperature the charge relaxation resistance vanishes in agreement with the recent zero temperature analysis [M. Filippone and C. Mora, Phys. Rev. B {\bf 86}, 125311 (2012) and P. Dutt, T. L. Schmidt, C. Mora, and K. Le Hur, Phys. Rev. B {\bf 87}, 155134 (2013)].
\end{abstract}
\date{\today}


\maketitle

%
\section{Introduction\label{Sec:Intro}}

\begin{figure}[b]
\centerline{\includegraphics[width=7cm]{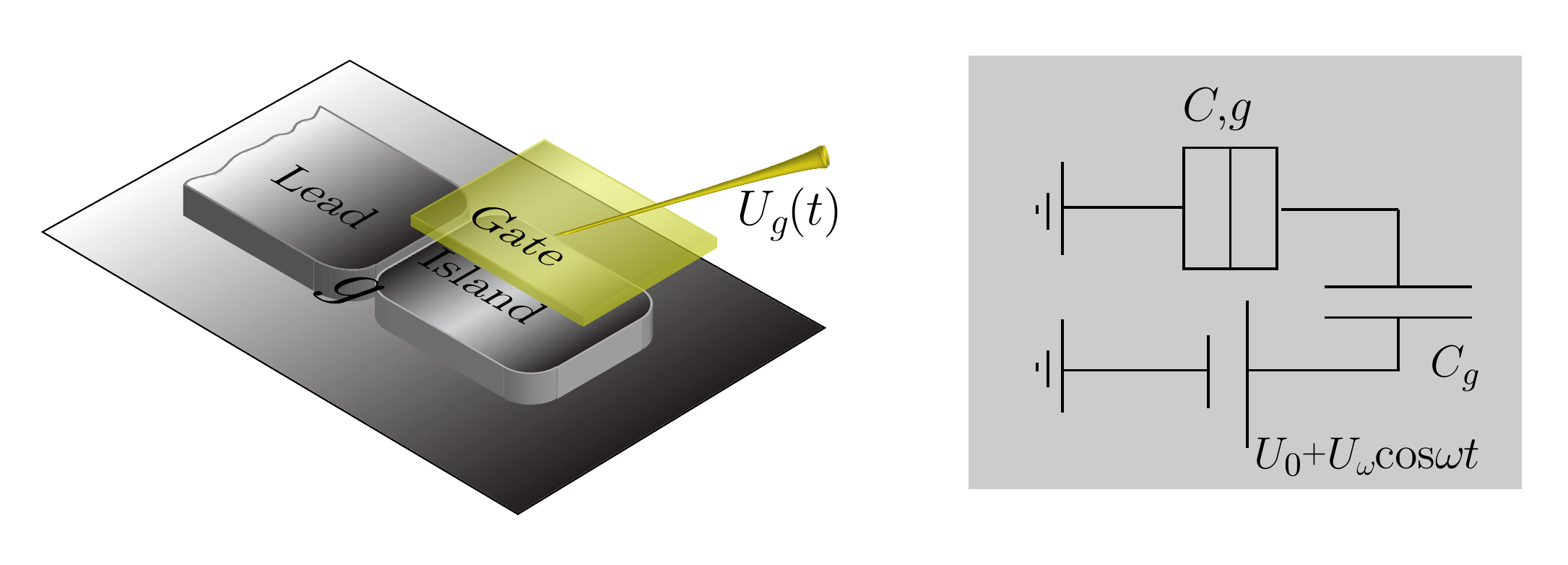}}
  \caption{(Color online) The set-up: a SEB subjected to a time-dependent gate voltage $U_g(t)$ (left) and
    the equivalent electric circuit (right).}
          \label{figure1}
\end{figure}

During the last decades Coulomb blockade has become a powerful tool for
observation of interaction and quantum effects in single electron devices [\onlinecite{zaikin,ZPhys,grabert,blanter,aleiner,Glazman}]. This phenomenon
is widely observed in low temperature electron transport through a single electron transistor.
Another system which low temperature properties are affected by Coulomb blockade is a single
electron box (SEB). It is schematically shown in Fig. \ref{figure1}. Small metallic island is
coupled capacitively to the gate electrode with the voltage $U_g$. The number of electrons on the island is not conserved due to the tunneling in and out of an equilibrium electron reservoir. A time dependent gate voltage $U_g(t)$ generates ac current through the device.

The equivalent electric circuit of a SEB  (see Fig. \ref{figure1}) is characterized by two capacitances. The gate
capacitance $C_g$ controls the external (induced) charge $q$ on the island, $q= C_g U_g$. The total capacitance $C$ determines the so-called charging energy $E_c=e^2/2C$. It is  the latter that is responsible for the Coulomb blockade effects. The tunnel junction is characterized by the dimensionless (in units $e^2/h$) conductance $g$. Throughout the paper we use a standard assumption that the Thouless energy of the island is the largest energy scale in the problem. This allows us to work in a zero dimensional approximation neglecting spatial dependence of all quantities.

Since there is no dc transport through the SEB, an essential dynamic characteristic becomes the admittance which characterizes the response of ac current $I_\omega$ to the infinitely small ac part $U_\omega$ of the time dependent gate voltage $U_g(t)=U_0+U_\omega \cos\omega t$: $\mathcal{G}(\omega) = I_\omega/U_\omega$. Long ago it was demonstrated that the admittance of SEB is affected by Coulomb blockade at low temperatures $T\ll E_c$ [\onlinecite{Nazarov1990}]. However, since then the majority of works have addressed the so-called quantum capacitance: $C_{\rm eff}=\partial Q/\partial U_g$, where $Q$ is the average charge on the island, which determines the imaginary part of the admittance~[\onlinecite{matveev,Grabert0a,Grabert0b,matveev1,Grabert1,Grabert2,beloborodov1}]. Classically, at high temperatures $T\gg E_c$ the effective capacitance coincides with $C_g$. As temperature decreases $C_{\rm eff}$ starts to deviate from $C_g$ due to interaction and coherence effects. In seminal paper [\onlinecite{buttiker0}] it was suggested that the real and imaginary parts of the admittance in a SEB can be related in an universal way. After paper [\onlinecite{buttiker0}] a SEB admittance have attracted significant theoretical interest [\onlinecite{buttiker3,buttiker2,buttiker1,imry,Park}]. The admittance in the quasi-static regime was measured in a single channel SEB constructed in 2D electron gas [\onlinecite{gabelli}]. At present, there exists a number of measurements of admittances for different realizations of a SEB performed with the help of radio-frequency reflectometry [\onlinecite{delsing,Ciccarelli,Chorley,Frake}].

The classical electrodynamics of a SEB suggests the following expression for the admittance at low frequencies, $\omega \ll g E_c$:
\begin{equation}
  \label{admittance_cl}
   \mathcal{G}(\omega)=-i \omega C_g+ \omega^2 C_g C R ,
\end{equation}
where $R= h/(e^2 g)$ stands for the classical resistance of the tunnel junction. In Ref. [\onlinecite{buttiker0}] the following generalization of the classical result \eqref{admittance_cl} has been proposed for the quantum coherent SEB with $C=C_g$:
 \begin{equation}
  \label{admittance_q}
   \mathcal{G}(\omega)=-i \omega C_{\rm eff}+ \omega^2 C_{\rm eff}^2 R_q ,
\end{equation}
where $R_q$ was termed as \emph{charge relaxation resistance}. Treating the Coulomb interaction within the Hartree-Fock approximation, the authors of Ref. [\onlinecite{buttiker0}] demonstrate that for single channel tunnel junction the charge relaxation resistance in Eq. \eqref{admittance_q} becomes universal, $R_q=h/(2e^2)$.
The full quantum mechanical treatment of the charging energy in the case of single channel  tunnel junction
demonstrates that at zero temperature $R_q=h/(2e^2)$ ($R_q=h/e^2$)  for frequencies $\omega \ll \delta$ ($\omega\gg\delta$) [\onlinecite{Mora2010}]. Here $\delta$ denotes the mean level spacing of single particle states inside the island of a SEB. Both results follow from two observations: (i) the effective low
energy Hamiltonian of single channel SEB is of Fermi liquid type; (ii) the Korringa-Shiba relation [\onlinecite{Korringa1950},\onlinecite{Shiba1975}] for the response function $i  \mathcal{G}(\omega)/\omega$ holds  within Fermi liquid low-energy description [\onlinecite{Mora2010}]. Recently, the analysis of Ref. [\onlinecite{Mora2010}] has been generalized to the case of a SEB with a weak ($g\ll 1$) multi channel tunnel junction and a large island, $\delta \to 0$. It was found [\onlinecite{Filippone2012},\onlinecite{Mora2013}] that at zero temperature the charge relaxation resistance is inversely proportional to the number of channels in a tunnel junction and is independent of the external charge; $R_q$ vanishes in the limit of infinite number of channels for any value of $q$.

At finite temperatures the SEB with a multi channel tunnel junction in the limit of negligible mean level spacing, $\delta \to 0$, has been analyzed  in Ref. [\onlinecite{Rodionov2009}]. In particular, it was demonstrated that in the limit of weak  tunneling, $g\ll 1$, and near the charge degeneracy points the SEB admittance at low frequencies ($\omega \ll g \max\{|\Delta|, T\}$) can be set down in the following form
\begin{equation}
  \label{admittance_r}
   \mathcal{G}(\omega)=-i \omega C_{\rm eff}+ \omega^2 \frac{C}{C_g} \mathcal{C}_g^2 \mathcal{R}_q .
\end{equation}
Here $\Delta$ denotes the electrostatic energy due to one excess electron on the SEB island. It depends on the external charge $q$ and  satisfy inequality $|\Delta|\ll E_c$ near a charge degeneracy point. The quantity $\mathcal{C}_g = \partial \mathcal{Q}/\partial U_0$ stands for the renormalized gate capacitance which measures the response of the effective charge $\mathcal{Q}$, introduced in Refs. [\onlinecite{Burmistrov2008},\onlinecite{Burmistrov2010}] by one of us, to the static part of the gate voltage.
Contrary to the average charge $Q$ on the island, the effective charge is expected to be integer quantized
at zero temperature [\onlinecite{Burmistrov2008,Burmistrov2010,Semenov2013}]. This implies that $\mathcal{C}_g$ vanishes at $T=0$ contrary to $C_{\rm eff}$.  The charge relaxation resistance in Eq. \eqref{admittance_r} is determined by the renormalized tunneling conductance $g(T)$, $\mathcal{R}_q = h/(e^2 g(T)) \gg h/e^2$. The very same conductance $g(T)$ determines the dc conductance of the single electron transistor (SET) under small bias between source and drain. We also note that Eq. \eqref{admittance_r} was proposed for the SEB with arbitrary relation between $C$ and $C_g$. For the case of weak tunneling the treatment of Ref. [\onlinecite{Rodionov2009}] was restricted to the sequential tunneling approximation dressed by the renormalization due to virtual processes. The processes of inelastic cotunneling [\onlinecite{Averin1990}] which dominate the dc transport through the SET at low temperatures $T\ll T_{\rm in} \sim |\Delta|/\ln(1/g)$ were not taken into account. Therefore, the extrapolation of Eq. \eqref{admittance_r} down to the zero temperature and comparison with the result of Ref. [\onlinecite{Filippone2012},\onlinecite{Mora2013}] were not possible.

The real part of the admittance of a SEB with a multi channel tunnel junction in the regime of inelastic cotunneling has been studied in Ref. [\onlinecite{Nazarov1990}]. It was found that the real part of admittance is proportional to $g^2 \omega^2 \max\{T^4,\omega^4\}/E_c^2$. This results implies the zero charge relaxation resistance at $T=0$ in agreement with the result of Ref. [\onlinecite{Filippone2012},\onlinecite{Mora2013}] extrapolated to limit of the infinite number of channels in the tunnel junction. However, the analysis of Ref. [\onlinecite{Nazarov1990}] has been restricted to Coulomb valleys, i.e. to integer values of $q$.

In this paper we address the following question: how the zero temperature result for the charge relaxation resistance obtained in Refs. [\onlinecite{Filippone2012},\onlinecite{Mora2013}] crosses over to the finite temperature result of Ref. [\onlinecite{Rodionov2009}]? To answer this question we performed a detailed study of the admittance of the multi channel SEB  near the charge neutrality points in the low temperature regime where the inelastic cotunneling processes dominate the dynamics. We found that the real part of admittance is proportional to $g^2 \omega^2\max\{T^2,\omega^2\}/\Delta^4$. The charge relaxation resistance (extracted from Eq. \eqref{admittance_q}) is strongly temperature dependent and small, $R_q\sim (h/e^2)(T/\Delta)^2 \ll h/e^2$. In agreement with Refs. [\onlinecite{Mora2010},\onlinecite{Filippone2012}], we obtained that $R_q$ is independent of $g$ and vanishes at zero temperature.  Our explicit results demonstrate strong violation of Korringa-Shiba relation for the response function $i  \mathcal{G}(\omega)/\omega$ and, consequently, support the non-Fermi liquid behavior of the multi channel SEB near the charge degeneracy points.

The structure of the paper is as follows. In Sec. \ref{Sec:Second} we introduce the Hamiltonian and Kubo formula for admittance of a single electron box. The pseudofermion representation for the low energy Hamiltonian valid in the cotunneling regime is presented in Sec. \ref{Sec3}. The results of calculation of the admittance at low frequencies to the second order in the tunneling conductance $g$ are given in Sec. \ref{Sec4}. Finally, discussion of our results and conclusions are presented in Sec. \ref{Sec:Conc}. The details of calculations are summarized in Appendix \ref{Sec:App}. We use units with $\hbar=e=1$ through out the paper except for the final results.

\section{Formalism\label{Sec:Second}}

\subsection{Hamiltonian}

We start with the standard Hamiltonian describing Coulomb blockade in a SEB [\onlinecite{Mezei1971,Shekhter1972,Kulik1975}]:
\begin{gather}
   \label{ham1}
      H=H_l + H_d + H_c + H_t .
\end{gather}
Here $H_l$ ($H_d$) denotes free electron Hamiltonian in the lead (the island),
\begin{equation}
   \label{ham2}
    \begin{split}
    H_l&=\sum_{k} \varepsilon^{(a)}_{k}a^\dagger_{k}a_{k} ,\\
    H_d & =
    \sum_\alpha\varepsilon^{(d)}_\alpha d^\dagger_{\alpha}d_{\alpha},
     \end{split}
\end{equation}
where operators $a^\dag_{k}$ ($d^\dag_{\alpha}$) create an electron
in the reservoir (the island). Energies $\varepsilon^{(a)}_k, \varepsilon^{(d)}_\alpha$ are counted from the chemical potential. The Hamiltonian
\begin{equation}
   \label{ham3}
      H_c=E_c\big ( \hat{n}_d-q\big)^2, \qquad  \hat{n}_d=\sum_{\alpha}d^\dagger_{\alpha}d_{\alpha},
\end{equation}
takes into account the electrostatic energy due to the finite size of the island.

The Hamiltonian
\begin{gather}
   \label{ham-tun}
    H_{t}=\sum_{k,\alpha}t_{k\alpha} a_{k}^\dagger d_{\alpha}+{\rm
    h.c.}
\end{gather}
describes tunneling of electrons between the island and the reservoir. In order to characterize the tunnel junction, following Ref. [\onlinecite{Rodionov2009}], we introduce Hermitian matrix:
\begin{equation}
   \hat{g}_{\alpha\alpha^\prime}=(2\pi)^2
   \left[\delta(\varepsilon^{(d)}_{\alpha})\delta(\varepsilon^{(d)}_{\alpha^\prime})
   \right]^{1/2}\sum_k t^\dagger_{\alpha
   k}
   \delta(\varepsilon^{(a)}_{k})t_{k\alpha^\prime}
\end{equation}
acting in the Hilbert space of the island's states. Here the delta-functions are assumed to be smoothed on
some intermediate scale between $\delta$ and $\min\{T, |\omega|\}$. The matrix $\hat{g}$ allows one to define the number of open channels $N_{\rm ch}$ and the effective channel conductance $g_{\rm ch}$:
\begin{equation}
   \label{aes-condition1}
N_{\rm ch}=\frac{(\tr\hat{g})^2}{\tr \hat{g}^2}, \qquad  g_{\rm ch}=\frac{\tr \hat{g}^2}{\tr \hat{g}}.
  \end{equation}
The dimensionless conductance $g$ which characterizes the tunnel junction in classical electrodynamics is given as $g=g_{\rm ch}N_{\rm ch}$.

  In the present paper we assume that $N_{\rm ch}\gg 1$ and $1/N_{\rm ch}^{2}\ll g_{\rm ch}\ll 1$. Although within these assumptions the classical conductance $g$ can be still large, in what follows we restrict our consideration to the case $1\gg g\gg 1/N_{\rm ch}$.
We are interested in temperatures much smaller than the charging energy but much larger than the mean level spacing, $E_c\gg T \gg \delta$.

\subsection{Admittance and polarization operator}

The admittance of a SEB being the linear response of the ac current to the ac part of the time-dependent gate voltage, $U_g(t)=U_0+U_\omega\cos \omega t$, can be expressed as [\onlinecite{Rodionov2009}]
\begin{gather}\label{admgen}
 \mathcal{G}(\omega)=-i\omega C_g\bigl (1+\Pi^R(\omega)/C\bigr ) ,
\end{gather}
where $\Pi^R(\omega)$ stands for the Fourier transform of the retarded polarization operator of electrons on the island:
\begin{equation}
  \label{polar-0}
 \Pi^R(t)=i \Theta(t) \bigl \langle[\hat n_d(t),\hat n_d(0)]\bigr \rangle .
\end{equation}
Here $\Theta(t)$ denotes the Heaviside step function. In the quasi-static regime $\omega\to 0$, the polarization operator $\Pi^R(\omega)$ can be expanded in regular series in $\omega$:
\begin{equation}
   \label{polar-3}
   \Pi^R(\omega)=\pi_0+i\omega\pi_1+{\cal O}(\omega^2) ,
\end{equation}
where both $\pi_0$ and $\pi_1$ are real functions. We stress that we assume $|\omega| \gg \delta$
throughout the paper. The static part $\pi_0$ is fully determined by the average charge on the
island $Q$:
\begin{equation}
\pi_0 = \frac{C}{C_g} C_{\rm eff} - C ,\label{SWId}
\end{equation}
where we remind $C_{\rm eff} = \partial Q/\partial U_0$ with $Q=\langle \hat n_d \rangle$.
This result holds by virtue of the Ward identity which relates the static polarization operator and the compressibility~[\onlinecite{AGD}]. We note that on the classical level $C_{\rm eff} = C_g$ and $\pi_0=0$.
The classical electrodynamics result \eqref{admittance_cl} implies $\pi_1=2\pi C^2/g$ on the classical level. In what follows we discuss how quantum effects due to inelastic cotunneling change naive classical expectations for $\pi_0$ and $\pi_1$.

\section{Weak tunneling regime \label{Sec3}}

\subsection{Projected Hamiltonian}

In what follows, contrary to Ref. [\onlinecite{Nazarov1990}], we confine our consideration to the vicinity of the one of charge degeneracy points, i.e. points where the external charge $q=k+1/2$ where $k$ is an integer. At these points the gap $\Delta = 2E_c (k+1/2-q)$ between the ground and the first excited state of the charging Hamiltonian $H_c$ vanishes. In the vicinity of the degeneracy point, the gap $\Delta$ is small in comparison with the charging energy, $|\Delta|\ll E_c$. The processes of inelastic cotunneling becomes important at low temperatures $T\ll |\Delta|$. At $|\Delta|\ll E_c$, one can truncate the Hilbert space of electrons on the isolated island to two charging states characterized by $Q=k$ and $Q=k+1$ [\onlinecite{matveev}]. The projected Hamiltonian acquires a
form of $2\times2$\ matrix acting in the isospin $1/2$ space of these two charging
states [\onlinecite{matveev}]:
\begin{equation}
  \label{ham4}
   \tilde{H} =H_l+H_d + \tilde{H}_t + \Delta S_z + \frac{\Delta^2}{4E_c}+\frac{E_c}{4} ,
\end{equation}
where $H_{l,d}$ are given by Eq.~\eqref{ham2}, and
\begin{equation}
 \tilde{H}_t=\sum_{k,\alpha}t_{k\alpha} a^\dagger_k d_\alpha
   S^++\hbox{h.c.} \, .
\end{equation}
Here $S^z,\ S^{\pm}=S^x\pm iS^y$\ stand for standard spin $1/2$ operators. The Hamiltonian \eqref{ham4} is the Hamiltonian for the $N_{\rm ch}$ channel Kondo model. The gap $\Delta$ between the charging states plays the role of an effective magnetic field.

In the presence of ac component of the gate voltage, the energy detuning from the degeneracy point becomes time dependent: $\Delta(t) = \Delta-
(C_g/C) U_\omega\cos\omega t$.  Then, as follows from Eq. \eqref{ham4}, the linear response of a SEB to
ac gate voltage $U_\omega$ is determined by the longitudinal dynamical isospin susceptibility:
\begin{equation}
  \label{spin-correlator}
  \Pi^R_s(t)=i \Theta(t) \bigl  \langle[S_z(t),S_z(0)] \bigr \rangle  .
\end{equation}
In particular, the admittance $\mathcal{G}(\omega)$ is given as [\onlinecite{Rodionov2009}]:
\begin{equation}
\label{spin-dissipation}
   \mathcal{G}(\omega)=-i\omega\frac{C_g}{C}\Pi_{s}^{R}(\omega) .
  \end{equation}
We note that the average charge on the island can be written as $Q=k+1/2- \langle S_z \rangle$, i.e. it is related to the isospin magnetization. As the consequence of Eq. \eqref{SWId}, the dynamical isospin susceptibility should satisfy the relation $\Pi_{s}^{R}(0)= {C} C_{\rm eff}/C_g$.

\subsection{Pseudofermion effective action}

To deal with spin operators we employ the method of Abrikosov's
pseudofermion operators. Following Ref. [\onlinecite{abrikosov}], we introduce
fermion operators $\bar{\psi}_\alpha$,
$\psi_\alpha$ such that
\begin{gather}
   \label{pseudofermion}
  {\bm S}= \frac{1}{2} \bar{\psi}_\alpha \bm{\sigma}_{\alpha\beta}\psi_\beta .
\end{gather}
Here $\bm{\sigma}=\{\sigma_x, \sigma_y, \sigma_z\}$ denotes standard  Pauli matrices.
To exclude redundant unphysical states (the states with
$\sum_\alpha \bar{\psi}_\alpha \psi_\alpha>1$) we add to the Hamiltonian $\tilde{H}$ an artificial chemical potential $\eta$ for the pseudofermions. It is necessary to take the limit $\eta\rightarrow -\infty$ at
the end of any calculation.

We remind that the physical partition function ${\cal Z}$ and correlation functions $\langle \mathcal{O}\rangle$ can be found from the pseudofermion ones with the help of the following rules:
\begin{gather}
 \label{pf-partition}
  \begin{split}
  {\cal Z}&=\lim_{\eta\rightarrow-\infty}\frac{\partial}{\partial e^{\beta\eta}}{\cal Z}_{\rm pf} ,\\
  \langle{\cal O}\rangle&=\lim_{\eta\rightarrow-\infty}\bigg\{ \langle{\cal O}\rangle_{\rm pf}+
  \frac{{\cal Z}_{\rm pf}}{{\cal Z}}\frac{\partial}{\partial e^{\beta\eta}}\langle{\cal O}\rangle_{\rm pf}\bigg\} .
  \end{split}
\end{gather}

In the case $N_{\rm ch} \gg 1$ and $1\gg g \gg 1/N_{\rm ch}$, after the integration over electrons in the reservoir and the island the Hamiltonian \eqref{ham4} can be transformed into the following imaginary-time effective
action [\onlinecite{Rodionov2009}]:
\begin{equation}
 \label{pf-action}
  \begin{split}
   S&= \frac{\beta \Delta^2}{4E_c}+
  \int_0^\beta d\tau\bar{\psi}\Big(\partial_\tau+\frac{\sigma_z \Delta}{2}-\eta\Big)\psi \\
  &+\frac{g}{4}\int_{0}^\beta d\tau_1d\tau_2\alpha(\tau_{12})[\bar{\psi}(\tau_1)\sigma_-\psi(\tau_1)]
   [\bar{\psi}(\tau_2)\sigma_+\psi(\tau_2)]
   .
  \end{split}
\end{equation}
Here $\sigma_{\pm}=(\sigma_x\pm i\sigma_y)/2$ and the kernel
\begin{equation}
\alpha(\tau)=\frac{T}{\pi}\sum\limits_{\omega_n} |\omega_n| e^{-i\omega_n \tau},
\end{equation}
 where $\omega_n=2\pi T n$ is the bosonic Matsubara frequency.
We note that the action similar to Eq.~\eqref{pf-action} has been first analyzed by Larkin and Melnikov
in Ref. [\onlinecite{larkin}]. Effective action~\eqref{pf-action} corresponds to the XY Bose-Kondo
model for the spin $1/2$ [\onlinecite{Sachdev},\onlinecite{Si},\onlinecite{Demler}].

The dynamical spin susceptibility \eqref{spin-correlator} is determined by the pseudofermion dynamical spin susceptibility:
\begin{equation}
  \label{polarization1}
  \Pi_{s,\textrm{pf}}(\tau)=\frac{1}{4}\langle{\cal T}_\tau[\bar{\psi}(\tau)\sigma_z\psi(\tau)][\bar{\psi}(0)\sigma_z\psi(0)]\rangle ,
\end{equation}
where ${\cal T}_\tau$ denotes the imaginary-time ordering and the average is taken with respect to the effective action~\eqref{pf-action}. According to Eq.~\eqref{pf-partition} we also need  the expression for the physical partition function. It can be written as
\begin{equation}
  \label{pf-partition1}
 {\cal Z} =\lim\limits_{\eta\to 0}{\cal Z}_{\textrm{pf}}e^{-\beta\eta}\sum_\sigma \mathcal{G}_\sigma(\tau)\big|_{\tau\rightarrow0^-} ,
\end{equation}
where $\mathcal{G}_\sigma(\tau) = -\langle {\cal T}_\tau
\psi_\sigma(\tau)\bar{\psi}_\sigma(0)\rangle$ stands for the exact
imaginary-time pseudofermion Green's function.

As we discussed in the Introduction, in the Fermi liquid the imaginary and real parts of the dynamical spin susceptibility are related by the so-called Korringa-Shiba relation [\onlinecite{Korringa1950},\onlinecite{Shiba1975}]. Taking into account that $\Delta$ in the effective action \eqref{pf-action} plays a role of magnetic field, Korring-Shiba relation for $\Pi^R_s$ should have the following universal form:
\begin{align}
  \Im \Pi^R_s(\omega) \overset{\displaystyle \large ?}{=} 2\pi \omega  \bigl [\Re \Pi_s^R(0)\bigr ]^2,\qquad  \omega \to 0.
  \label{eq:KS}
\end{align}
If this relation were correct, the low-frequency admittance $\mathcal{G}(\omega)=-i\omega ({C_g}/{C}) ( C_{0} + i \omega C_0^2 R_q )$ would be characterized by the universal charge relaxation resistance, $R_q = h/e^2$, similar to the single channel case.
Here we introduce $C_0 = (C/C_g) C_{\rm eff}$. As we shall demonstrate below by direct calculation, Korringa-Shiba relation \eqref{eq:KS} \emph{does not hold} for the effective action \eqref{pf-action}.

\begin{figure}[t]
\centerline{ \includegraphics[width=70mm]{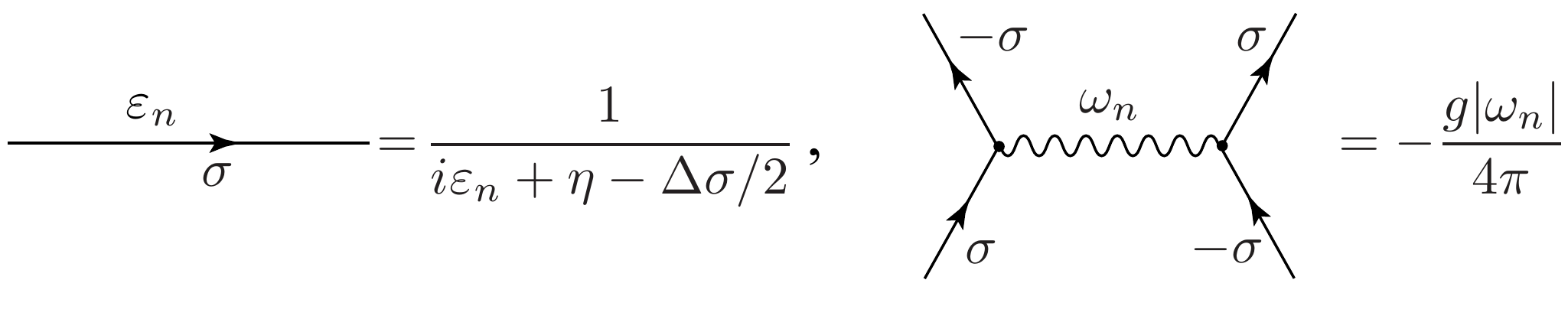}}
   \caption{Feynman rules for the pseudofermion action \protect \eqref{pf-action}.}
   \label{figure2}
\end{figure}

\section{Admittance in the cotunneling regime \label{Sec4}}

The effective action \eqref{pf-action} is suitable for the perturbation theory in $g\ll 1$. Here we evaluate the imaginary part of the dynamical spin susceptibility $ \Pi_{s}(\omega)$ to the second order in $g$.  The Feynman rules for action \eqref{pf-action} are shown in Fig.~\ref{figure2}. The bare pseudofermion Matsubara Green's function is given as follows:
\begin{equation} \label{ZerothOrderg}
G_\sigma(i\varepsilon_n) =\frac{1}{i\varepsilon_n +\eta - \sigma
\Delta/2}.
\end{equation}

Thus from Eqs. \eqref{polarization1}  and \eqref{pf-partition1} in the zeroth order in $g$ we find
\begin{equation}
{\cal Z}^{(0)}_{\rm pf}=1, \quad {\cal Z}^{(0)} =2\cosh\frac{\beta\Delta}{2} ,
\quad \Pi^{(0)}_{s,\textrm{pf}}(i\omega_n)=0\, .
\label{eq:zero:order}
\end{equation}

\subsection{Perturbation theory in $g$: Sequential tunneling}

\begin{figure}[t]
  \includegraphics[width=70mm]{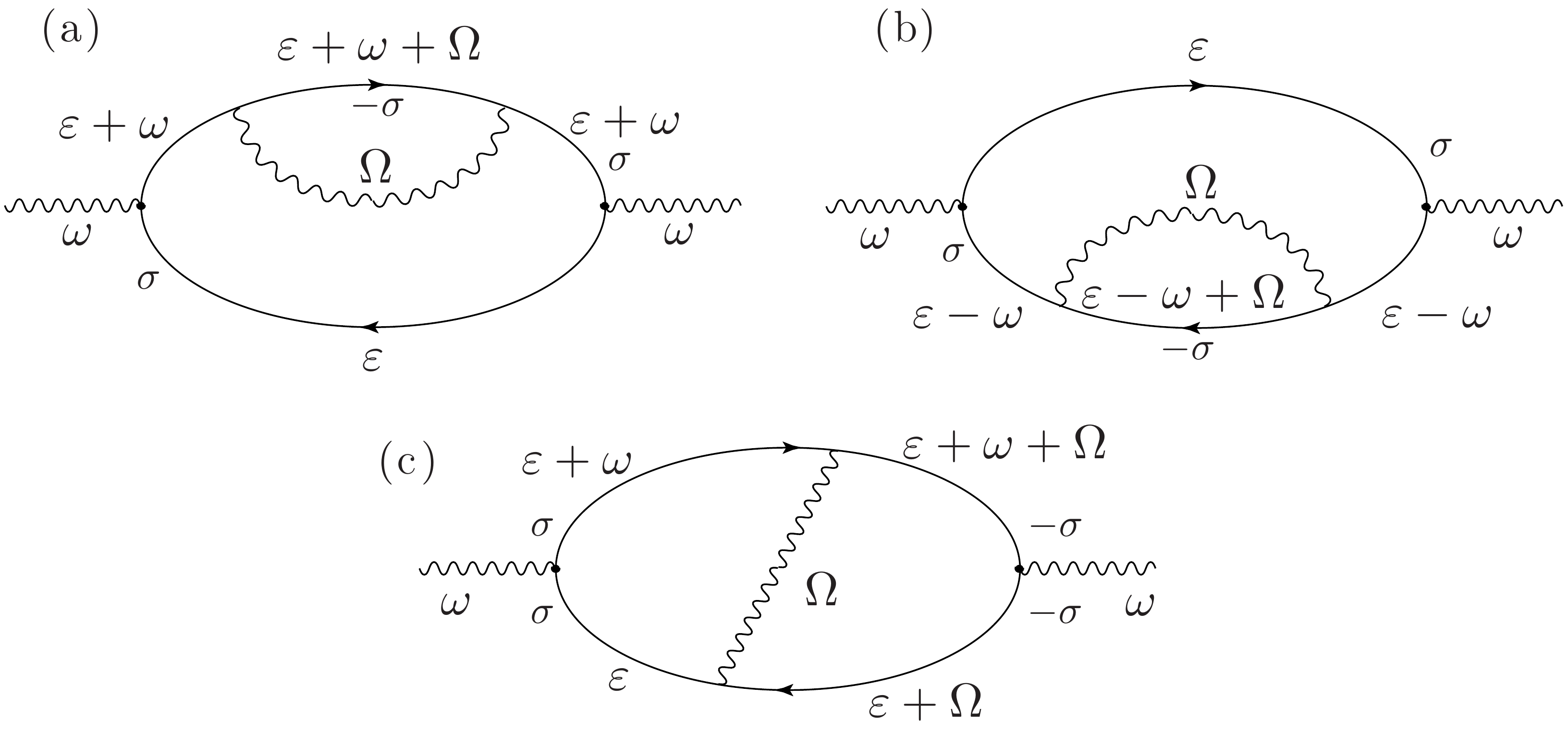}
   \caption{Feynman diagrams for the pseudofermion dynamical spin susceptibility in the first order in $g$.}
\label{figure3}
\end{figure}

Before discussing the inelastic cotunneling regime (second order in $g$) we remind briefly the result of Ref. [\onlinecite{Rodionov2009}] for the pseudofermion dynamical spin susceptibility in the regime of sequential tunneling. In Fig. \ref{figure3} we present diagrams contributing to the pseudofermion dynamical spin susceptibility $\Pi_{s,\textrm{pf}}(i\omega_n)$
in the first order in $g$. Their evaluation demonstrates that  $\Im \Pi^R_{s}(\omega)$ suffers from  singularity at $\omega\to 0$:
\begin{equation}
\Im \Pi^{R,(1)}_{s}(\omega) = \frac{g}{4\pi \omega}\frac{\beta \Delta}{\sinh(\beta\Delta)} .
\label{eq:1o:g}
\end{equation}
This unphysical divergence stems from non-commutativity of the limits $\omega \to 0$ and $g\to 0$ in the structure of the $\Im \Pi^{R,(1)}_{s}(\omega)$.  Summing the ladder-type diagrams and taking into account finite (the lowest order in $g$) broadening of the pseudofermion Green's function, we obtain the following expression [\onlinecite{Rodionov2009}]:
\begin{equation}
\Pi^{R, (\textrm{lad})}_{s}(\omega) = \frac{g}{4\pi }\frac{\beta \Delta}{\sinh(\beta\Delta)} \left (-i \omega+\frac{g \Delta}{2\pi} \coth \frac{\beta\Delta}{2}\right )^{-1}.
\label{eq:lad}
\end{equation}
We note that the broadening of the singularity at $\omega=0$ in  $\Im \Pi^{R}_{s,\textrm{pf}}(\omega)$ is determined by the sum of in- and out-rate of single electron tunneling [\onlinecite{grabert}].
At low temperatures $T\ll |\Delta|$, the result \eqref{eq:lad} implies ($|\omega|\ll {g |\Delta|}/{2\pi}$):
\begin{equation}
\Im \Pi^{R, (\textrm{lad})}_{s}(\omega) = \frac{2\pi \beta \omega}{g |\Delta|} e^{- \beta|\Delta|} .
\label{eq:PST}
\end{equation}
Such Arrhenius-type dependence, $\exp(-|\Delta|/T)$, is characteristic of real processes in which an additional electron or hole remains on the island after each tunneling event. We remind that the SET conductance in the sequential tunneling approximation is also of the same Arrhenius-type form at low temperatures $T\ll |\Delta|$ [\onlinecite{grabert},\onlinecite{Rodionov2009}]. As it is well-known, at low temperatures SET conductance has also a power-law (in $T/|\Delta|$) contribution of the second order in $g$ due to the processes of inelastic cotunneling [\onlinecite{Averin1990}]. As we shall demonstrate in the next section there is a similar contribution to the admittance.

\begin{figure}[t]
  \centering
  \includegraphics[width=70mm]{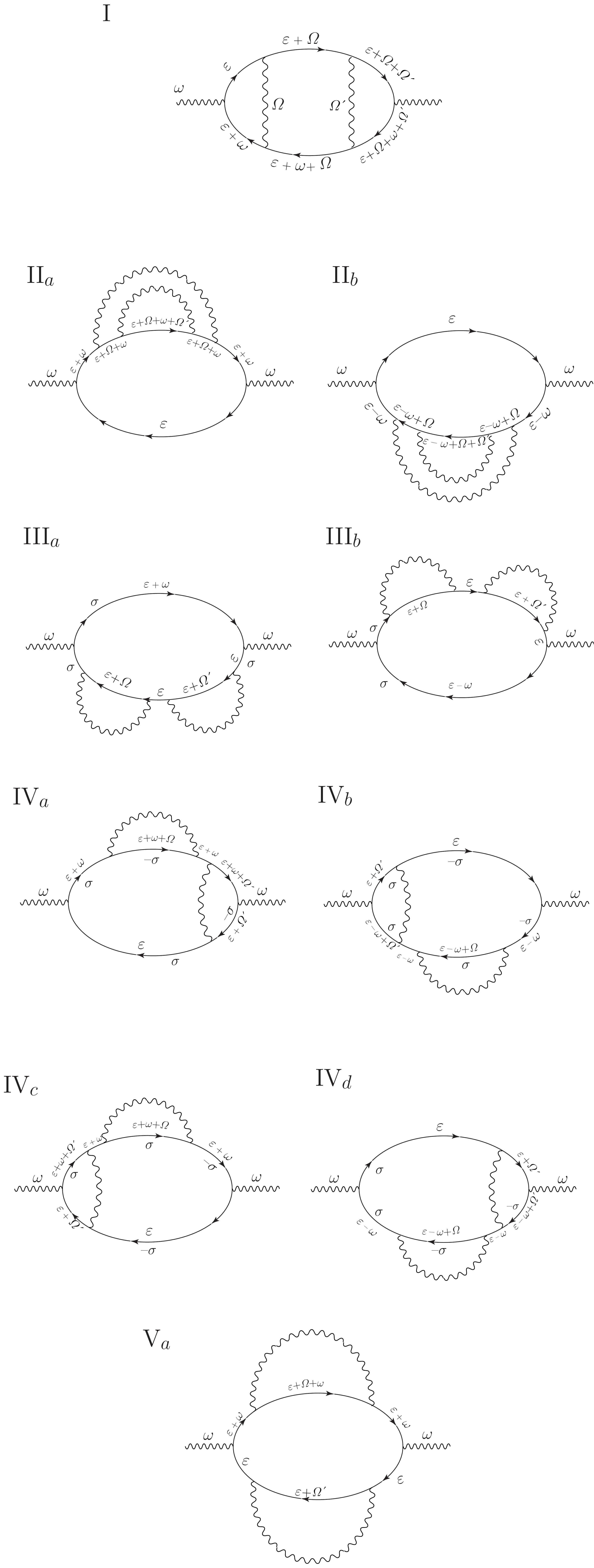}
  \caption{Second order in $g$ diagrams for the pseudofermion dynamical spin susceptibility.}
  \label{figure4}
\end{figure}

\subsection{Perturbation theory in $g$: Inelastic cotunneling}

The processes of inelastic cotunneling becomes important at temperatures $T\ll T_{\rm in}\ll |\Delta|$. However, these processes are of the second order in the tunneling conductance. Diagrams of the second order in $g$ for the pseudofermion dynamical spin susceptibility are shown in Fig. \ref{figure4}.  We remind that diagrams with pseudofermion loops vanish in the limit $\eta\rightarrow-\infty$. Evaluation of the ten diagrams in Fig. \ref{figure4} yields the following result for the imaginary part of the low-frequency dynamical spin susceptibility (see Appendix \ref{Sec:App})
\begin{gather}
   \Im\Pi^{R,(2)}_{s}(\omega)=\frac{g^2\omega}{24\pi\Delta^4} \left (T^2 +\frac{\omega^2}{4\pi^2}\right ), \quad |\omega|, T\ll |\Delta| .
\label{eq:PRR:2}
\end{gather}
This expression dominates over the contribution \eqref{eq:PST} due to sequential tunneling at low temperatures, $T\ll  T_{\rm in}$.

In the discussion above we do not consider renormalization of the effective action \eqref{pf-action} between the ultra violet energy scale of the order of $E_c$ and the infrared scale of the order of $\max\{|\Delta|,T\}$ [\onlinecite{larkin},\onlinecite{matveev},\onlinecite{schoeller}]. The renormalized effective action can be obtained from Eq. \eqref{pf-action} by the following substitutions $\psi \to \sqrt{Z} \psi$, $\bar{\psi} \to \sqrt{Z} \bar{\psi}$, $g \to \bar{g}= Z^2 g$ and $\Delta
\to \bar{\Delta}=Z^2 \Delta$, where the field renormalization factor $Z$ is given within one-loop approximation as [\onlinecite{larkin}]
\begin{equation}
Z=\left [1+\frac{g}{2\pi^2} \ln \frac{E_c}{\max\{|
\bar{\Delta}|,T\}}\right ]^{-1/2} .
\label{eq:Zren}
\end{equation}
We note that the pseudospin operator $S_z$ renormalizes according to $S_z \to Z^2 S_z$ [\onlinecite{Rodionov2009}]. Then from Eq. \eqref{eq:PRR:2} we find
\begin{gather}
   \Im\Pi^{R,(2)}_{s}(\omega)=\frac{Z^4 \bar{g}^2\omega}{24\pi\bar{\Delta}^4} \left (T^2 +\frac{\omega^2}{4\pi^2}\right )  .
\label{eq:PRR:2r}
\end{gather}
Remarkably, the field renormalization parameter $Z$ drops out from Eq. \eqref{eq:PRR:2r} such that it coincides with Eq. \eqref{eq:PRR:2} in spite of the renormalization.

Taking into account one-loop renormalization of the effective action, one finds the following result for the average charge on the island [\onlinecite{matveev}]:
\begin{equation}
\label{Qpf}
Q=k+\frac{1}{2}-\frac{Z^2}{2} \tanh\frac{\bar{\Delta}}{2T} .
\end{equation}
In the case of low temperatures, $T\ll T_{\rm in}$, Eq. \eqref{Qpf} can be simplified, and we obtain
\begin{equation}
\Re \Pi^{R,(1)}_{s}(0) = - \frac{\partial Q}{\partial \Delta} = \frac{Z^4 \bar{g}}{4\pi^2 |\bar{\Delta}|} .
\label{eq:RePi}
\end{equation}
We note that the factor $Z^4$ can be derived in the following way. Equation \eqref{eq:Zren} determines the one-loop renormalization group equation for the field renormalization factor. Then taking into account that (i) the relation $\partial \bar{\Delta}/\partial \Delta = Z^2$ holds within logarithmic accuracy and (ii) the renormalized conductance $\bar{g}$ corresponds to the energy scale $\bar{\Delta}$, one can obtain the result
\eqref{eq:RePi}.  Alternatively, the appearance of the factor $Z^4$ can be checked with the help of the expression for $Q$ derived to the second order in $g$ [\onlinecite{Burmistrov2010}].  We emphasize that Eq. \eqref{eq:RePi} implies that at $T\ll T_{\rm in}$ the effective capacitance becomes very different from $C_g$: $C_{\rm eff} = C_g Z^4 \bar{g} E_c/(2\pi^2 |\bar{\Delta}|)$.

Combining together Eqs. \eqref{eq:PRR:2} and \eqref{eq:RePi} we obtain the following result for the admittance of a SEB ($|\omega|  \ll T \ll T_{\rm in}$):
\begin{equation}
\mathcal{G}(\omega) =  - i \omega C_g \frac{Z^4 \bar{g} E_c}{2\pi^2|\bar{\Delta}|} +  \omega^2 C_g  \frac{Z^4 \bar{g}^2 T^2 E_c}{12\pi \bar{\Delta}^4} .
\label{eq:final:adm}
\end{equation}
This is the main result of the present paper. Results \eqref{eq:PRR:2r} and \eqref{eq:RePi} for the imaginary and real part of the dynamical isospin susceptibility implies \emph{strong violation} of the Korringa-Shiba relation \eqref{eq:KS} at low temperatures when the processes of inelastic cotunneling dominate. Using the Korringa-Shiba relation, one overestimates erroneously $\Im\Pi^{R}_{s}(\omega)$ by a large factor $(\bar{\Delta}/T)^2\gg 1$. The violation of the Korringa-Shiba relation signals that the Hamiltonian \eqref{ham4} and, consequently, the effective action \eqref{pf-action}, involves non-Fermi liquid physics.

\section{Discussion and conclusions \label{Sec:Conc}}

According to Eq. \eqref{admittance_q}, the result \eqref{eq:final:adm} for the admittance implies the following result for the charge relaxation resistance:
\begin{equation}
R_q = \frac{h}{e^2} \frac{\pi^2}{3} \left ( \frac{T}{ Z^2\bar{\Delta}}\right )^2, \qquad T  \ll T_{\rm in} \sim \frac{|\bar{\Delta}|}{\ln(1/\bar{g})} .
\label{eq:final:Rq}
\end{equation}
We emphasize that due to Coulomb interaction the charge relaxation resistance is strongly temperature and gate voltage dependent. Moreover, $R_q$ depends on the tunneling conductance $g$ although through the field renormalization factor only. Therefore, the charge relaxation resistance depends in nontrivial
way on the parameters of a SEB in contrast to the zero temperature predictions of Refs. [\onlinecite{Filippone2012},\onlinecite{Mora2013}] and original ideas of Ref. [\onlinecite{buttiker0}].
Also the charge relaxation resistance is much smaller than the resistance quantum, $R_q\ll h/e^2$. At $T=0$ the charge relaxation resistance vanishes, $R_q=0$. This behavior is in agreement with the zero temperature result of Refs. [\onlinecite{Filippone2012},\onlinecite{Mora2013}], $R_q(T=0)=(h/e^2)/N_{\rm ch}$, which implies zero charge relaxation resistance at $T=0$ and $N_{\rm ch} \to \infty$.

The real part of the admittance determines the energy dissipation rate of a SEB,
$\mathcal{W} = C_g E_c \Re \mathcal{G}(\omega) U_\omega^2$,  which appears due to the time dependent periodic gate voltage modulations. The result \eqref{eq:final:adm} leads to the ohmic dissipation at low frequencies:
\begin{equation}
\mathcal{W} =  \omega^2 \frac{Z^4 \bar{g}^2 T^2 E^2_c}{3\pi^2\bar{\Delta}^4} C^2_g  U_\omega^2, \quad |\omega|  \ll T \ll T_{\rm in} .
\label{eq:final:W}
\end{equation}
\begin{figure}[t]
  \centering
  \includegraphics[width=70mm]{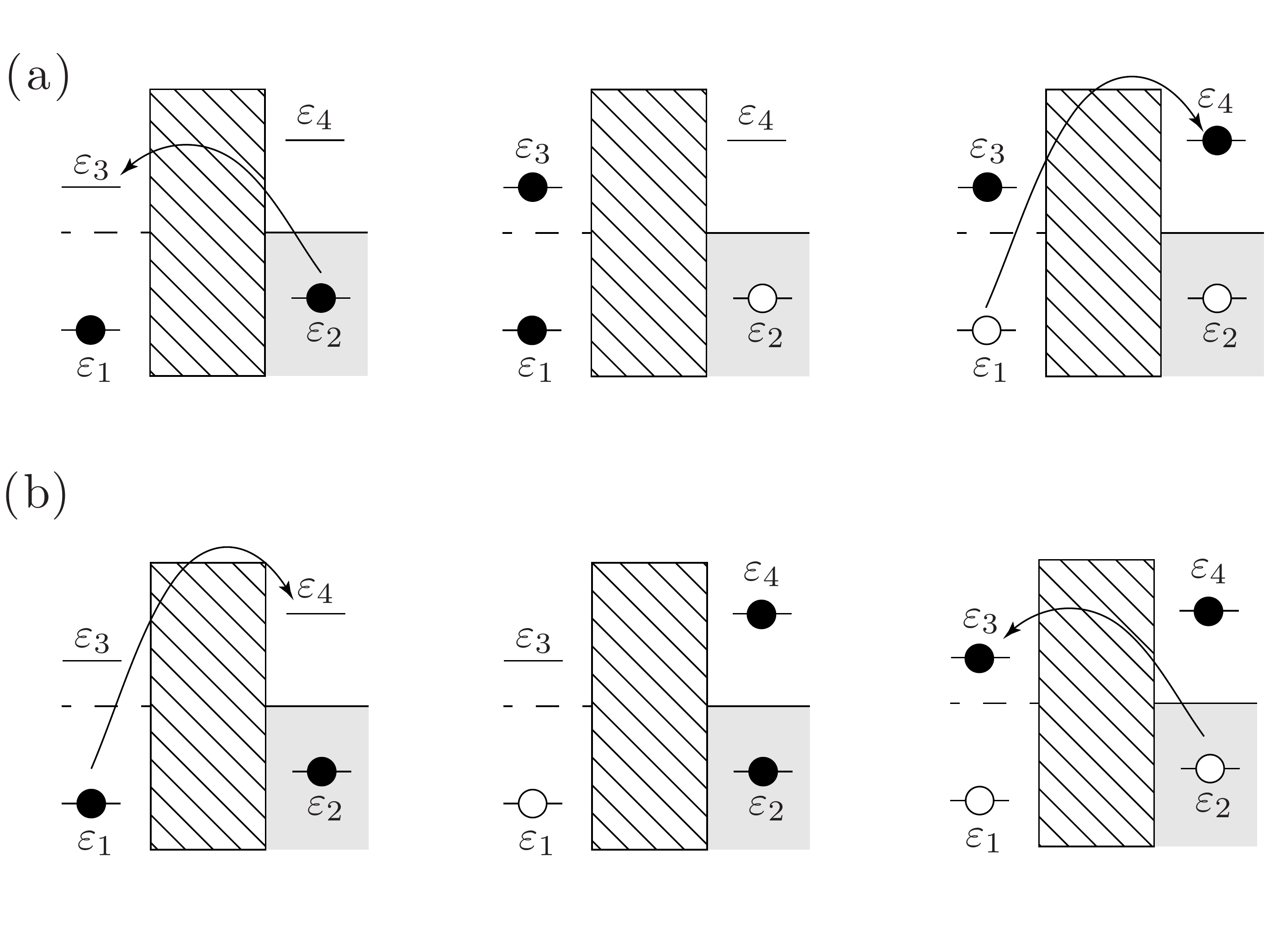}
  \caption{The processes of inelastic cotunneling. The intermediate state has an additional electron (a) and hole (b).}
  \label{figure5}
\end{figure}

This result for the dissipation rate has the following physical explanation. Let us estimate the rate $\Gamma_{\rm in}$ for the two-electron process in which one electron with energy $\varepsilon_1$ tunnels from the island into the reservoir and occupies the state with energy $\varepsilon_4$ whereas the other electron with energy $\varepsilon_2$ tunnels from the reservoir into the island occupying the state with energy $\varepsilon_3$ (see Fig. \ref{figure5}). For $\varepsilon_1 \neq \varepsilon_3$ this process is inelastic and
results in the electron-hole pair on the island at the end. It can go through two different
 intermediate states: with an additional electron and an additional hole on the island. The former costs the energy of the order of $2E_c$ whereas the latter costs the energy of the order of $|\Delta|$. In the considered case $|\Delta|\ll E_c$, we can neglect the contribution due to the intermediate state with an additional electron. Provided such transition is accompanied by a periodic perturbation which supplies the energy $\omega$ to the final state, we can estimate the corresponding rate as follows
\begin{gather}
\Gamma_{\rm in} \sim \frac{g^2}{\Delta^2} \left (\prod_{j=1}^4 \int\limits_{-\infty}^\infty d\varepsilon_j \right ) f_F(\varepsilon_1) \bigl [1-f_F(\varepsilon_3) \bigr ]  f_F(\varepsilon_2) \notag \\
\times \bigl [1-f_F(\varepsilon_4) \bigr ] \delta(\varepsilon_3+\varepsilon_4-\varepsilon_1-\varepsilon_2-\omega) .
\label{eq:est:rate}
\end{gather}
Here we use the fact that typical electron or hole energies in the integral in Eq. \eqref{eq:est:rate} are of the order of $\max\{T,|\omega|\} \ll |\Delta|$. Then for $|\omega|\ll T\ll |\Delta|$ we find that the frequency dependent part of the rate is estimated as $\Gamma^{(\omega)}_{\rm in} \sim g^2 T^2\omega/\Delta^2$. In fact, this rate is similar to the rate derived for a SET biased by voltage in which case the role of $\omega$ is played by dc voltage [\onlinecite{Averin1990}]. The quadratic dependence of $\Gamma^{(\omega)}_{\rm in}$ on $T$ is responsible for the $T^2$ factor in the expression \eqref{eq:final:W} for the energy dissipation rate. Indeed, averaging $\Gamma^{(\omega)}_{\rm in}$ over time-dependent gate voltage, taking contribution proportional to $U^2_\omega$ and multiplying the result by $\omega$, one gets the estimate for the rate of energy dissipation, $\mathcal{W} \sim \omega ({\partial^2 \Gamma^{(\omega)}_{\rm in}}/{\partial \Delta^2}) (C_g U_\omega/C)^2$, which, up to numerical factors, coincides with the result \eqref{eq:final:W}. At zero temperature the rate of inelastic cotunneling is given as $\Gamma_{\rm in} \sim g^2 \omega^3/\Delta^2$. This suggests the following estimate for the energy dissipation rate $\mathcal{W} \sim \omega^4 g^2 E_c^2 C_g U_\omega/\Delta^4$, i.e. non-ohmic dissipation of energy at zero temperature. Using the result \eqref{eq:PRR:2r} with $T=0$, we obtain the following expression for the energy dissipation rate at zero temperature:
\begin{equation}
\mathcal{W} =  \frac{Z^4\bar{g}^2 \omega^4 E^2_c}{12\pi^4\bar{\Delta}^4} C^2_g  U_\omega^2, \quad  T\ll |\omega| \ll  |\bar{\Delta}| .
\label{eq:final:W2}
\end{equation}

It is worthwhile to discuss the result of Ref. [\onlinecite{Nazarov1990}] which is complementary to the result
of the present paper. In Ref. [\onlinecite{Nazarov1990}] the real part of admittance was analyzed in the regime of inelastic cotunneling but precisely at $q=k$, i.e. at Coulomb valleys. It was found that  $\Re \mathcal{G}(\omega) \propto g^2 \omega^2 \max\{T^4,\omega^4\}/E_c^6$. We emphasize that the result of
Ref. [\onlinecite{Nazarov1990}] is quite unexpected. For example, the contribution due to inelastic cotunneling is proportional to $T^2/\Delta^2$ near the charge neutrality points and to $T^2/E_c^2$ at Coulomb valleys [\onlinecite{Averin1990}]. Therefore, on the basis of our result one could expect that at Coulomb valleys the energy dissipation rate is given by Eqs. \eqref{eq:final:W} and \eqref{eq:final:W2}
with $\Delta$ substituted by $E_c$. However, this na\"ive argument leads to overestimation of the energy dissipation rate by large factor $E_c^2/\max\{T^2,\omega^2\}$. The result of Ref. [\onlinecite{Nazarov1990}]
comes from particular cancellation of terms proportional to $\omega^2 T^2/E_c^4$ and $\omega^4/E_c^4$ in $\Re \mathcal{G}(\omega)$ (see comment at the end of Appendix \ref{Sec:App}).

We mention that our result \eqref{eq:final:adm} is at odds with the expression \eqref{admittance_r}
proposed by us in Ref. [\onlinecite{Rodionov2009}]. Since the effective charge $\mathcal{Q}$ is expected to be robustly integer quantized at zero temperature [\onlinecite{Burmistrov2008,Burmistrov2010,Semenov2013}], the renormalized gate capacitance $\mathcal{C}_g$ is exponentially small at $T\ll |\bar{\Delta}|$ [\onlinecite{Burmistrov2010}]. The renormalized conductance is known to be proportional to the temperature squared, $g(T) \sim \bar{g}^2 T^2/\bar{\Delta}^2$, in the regime of the inelastic cotunneling [\onlinecite{Averin1990},\onlinecite{Burmistrov2010}]. Therefore, Eq.\eqref{eq:final:adm} suggests the exponential suppression of the energy dissipation at $T\ll |\bar{\Delta}|$ contrary to the result \eqref{eq:final:W}. Thus for $g\ll 1$ the expression \eqref{admittance_r} works within the sequential tunneling approximation dressed by renormalization due to virtual processes only.

The effective action \eqref{pf-action} predicts zero value of $\bar{g}$ under the renormalization in the infra red. The $N_{\rm ch}$ channel Kondo model  \eqref{ham4} has the unstable fixed point at finite value of $\bar{g}=g_* \sim 1/N_{\rm ch}$ [\onlinecite{matveev},\onlinecite{zarand}]. Therefore, our results obtained within the effective action \eqref{pf-action}  are applicable for the Hamiltonian \eqref{ham4} while $\bar{g} \gg g_*$. As follows from Eq. \eqref{eq:Zren},  this condition implies that our results hold not too close to the charge degeneracy point, $|\Delta|\gg (g E_c/\pi^2) \exp (-\pi^2/g_*)$. Since $g_*\sim 1/N_{\rm ch}$, the scale $(g E_c/\pi^2) \exp (-\pi^2/g_*)$ becomes extremely small already for not too very large values of $N_{\rm ch}$. Comparing our result \eqref{eq:final:Rq} with the zero temperature result of Refs. [\onlinecite{Filippone2012},\onlinecite{Mora2013}], we find that for the case of finite number of channels
the charge relaxation resistance is given by  Eq. \eqref{eq:final:Rq} for temperatures $T \gg |\bar{\Delta}|/\sqrt{N_{\rm ch}}$.

To summarize, we have studied the low frequency admittance of a multi channel SEB under a slowly oscillating gate voltage. Focusing on the regime of inelastic cotunneling, we have calculated the admittance $\mathcal{G}(\omega)$ (see Eq. \eqref{eq:final:adm}) near the charge degeneracy points. We found the following:

(i) At finite temperatures but low frequencies, $T_{\rm in}\gg T\gg|\omega|$, the energy dissipation rate (determined by the real part of the admittance) is ohmic and scales as the temperature squared, see Eq. \eqref{eq:final:W}, in agreement with qualitative arguments.

(ii) At zero temperature the energy dissipation rate is super-ohmic, $\sim \omega^4$, see Eq. \eqref{eq:final:W2},  in agreement with qualitative estimates.

(iii) The imaginary and real parts of the response function $i\mathcal{G}(\omega)/\omega$ do not satisfy Korringa-Shiba relation. This supports  the non-Fermi liquid behavior of the model near the charge degeneracy points.

(iv) The charge relaxation resistance $R_q$ is strongly temperature dependent and small, $R_q \sim(h/e^2) (T/\bar{\Delta})^2\ll h/e^2$. It vanishes at $T=0$ in agreement with the recent zero temperature analysis of Refs. [\onlinecite{Filippone2012},\onlinecite{Mora2013}].

(v) The relation between the real part of admittance and the effective charge $\mathcal{Q}$ conjected by us in Ref. [\onlinecite{Rodionov2009}] does not hold beyond the sequential tunneling approximation dressed by renormalization due to virtual processes.

Finally, we mention that our result \eqref{eq:final:adm} for the admittance can be tested in a single electron box with small metallic island via radio-frequency reflectometry measurements [\onlinecite{delsing},\onlinecite{Frake}]. Also we mention that following approach of Refs. [\onlinecite{Rodionov2010a},\onlinecite{Rodionov2010b}] our results can be extended to non-equlibrium conditions, e.g. different temperatures of the island and the reservoir [\onlinecite{Future}].

\begin{acknowledgments}
We acknowledge useful discussions with A. Ioselevich, Yu. Makhlin, and, especially, with C. Mora and Yu. Nazarov. The research was funded by Russian Science Foundation under the grant No. 14-02-00898.
\end{acknowledgments}

\appendix
\section{Computation of the polarization operator: Diagrams of the second order in $g$\label{Sec:App}}

In this appendix we present details of computation of the polarization operator within the second order perturbation theory in $g$. There are contributions from the ten diagrams shown in Fig.~\ref{figure4}.
The task is simplified considerably by the fact that we only need the imaginary part of the retarded polarization operator. Each diagram consists of six pseudofermion Green's function lines and two interaction lines. Thus each diagram involves the summation over three internal energies: fermionic $\varepsilon$ and two bosonic ones $\Omega, \Omega^\prime$. As usual the fermionic sum is easily undertaken with the help of the following identity: $T \sum_{\varepsilon} f(\varepsilon) = (4\pi i)^{-1}\oint d\varepsilon \tanh(\varepsilon/2T) f(\varepsilon)$ where the contour of integration circles around all the poles of $\tanh$.

\subsection{The diagram I}

The contribution from the diagram I to the polarization operator can be  written as
\begin{gather}
\Pi_{s,{\rm pf}}^{(2),{\rm I}}(i\omega_n) = - \frac{T^3 }{4} \sum_{\sigma=\pm} \sigma^2 \sum_{\varepsilon,\Omega,\Omega^\prime} \tilde{\alpha}(i\Omega)\tilde{\alpha}(i\Omega^\prime) G_\sigma(i\varepsilon)  \notag \\
\times
G_\sigma(i\varepsilon+i\omega_n)
G_{-\sigma}(i\varepsilon+i\Omega)G_{-\sigma}(i\varepsilon+i\Omega+i\omega_n)
\notag \\
\times
G_{\sigma}(i\varepsilon+i\Omega+i\Omega^\prime) G_{\sigma}(i\varepsilon+i\Omega+i\Omega^\prime+i\omega_n) ,
 \end{gather}
where we introduce the kernel  $\tilde{\alpha}(i\Omega) = g \alpha(i\Omega)/4 =  g|\Omega|/(4\pi)$. Evaluating the sum over the fermionic energy $\varepsilon$ and performing analytic continuation, $i\omega_n\to \omega+i0$, we obtain
\begin{gather}
\Im \Pi_{s,{\rm pf}}^{R,(2),{\rm I}}(\omega) =   - \frac{T}{4 \omega}  \sum\limits_{\sigma=\pm}\bigl ( f^\prime_\sigma + f^\prime_{-\sigma}\bigr )
\Im K^{R,(2,1)}_{\sigma}(\omega)
\notag \\
 +  \sum_{\sigma=\pm} \frac{f_\sigma-f_{-\sigma}}{2\omega} \Im \Bigl[ K^{R,(1,1)}_\sigma(\omega) \Bigr ]^2
 .
 \label{eq:A:d1:1}
\end{gather}
Here $f_\sigma = f_F(-\eta + \Delta \sigma/2)$, $f^\prime_\sigma = \partial f_F(\varepsilon)/\partial \varepsilon \big |_{\varepsilon=-\eta + \Delta \sigma/2}$ and  $f_F(\varepsilon) =  1/[1+\exp(\varepsilon/T)]$ denotes the Fermi-Dirac distribution function. The functions $K^{R,(n,m)}_{\sigma}(\omega)$ are retarded function corresponding to the following Matsubara function
\begin{equation}
K^{(n,m)}_{\sigma}(i\omega)  = T\sum_\Omega \frac{[\tilde{\alpha}(i\Omega+i\omega)-\tilde{\alpha}(i\Omega)]^n}{(i\omega)^{n}(i\Omega+\Delta\sigma)^m} .
\label{eq:A:def:K}
\end{equation}

\subsection{The diagrams IIa and IIb}

The contribution from the diagram IIa to the polarization operator can be  written as
\begin{gather}
\Pi_{s,{\rm pf}}^{(2),{\rm IIa}}(i\omega_n) = - \frac{T^3 }{4} \sum_{\sigma=\pm} \sigma^2 \sum_{\varepsilon,\Omega,\Omega^\prime} \tilde{\alpha}(i\Omega)\tilde{\alpha}(i\Omega^\prime) G^2_\sigma(i\varepsilon)  \notag \\
\times
G_\sigma(i\varepsilon+i\omega_n)
G^2_{-\sigma}(i\varepsilon+i\Omega)
G_{\sigma}(i\varepsilon+i\Omega+i\Omega^\prime) .
 \end{gather}
The contribution from the diagram IIb can be found from the expression above by reverting the sign of $\omega_n$: $\Pi_{s,{\rm pf}}^{(2),{\rm IIb}}(i\omega_n)=\Pi_{s,{\rm pf}}^{(2),{\rm IIa}}(-i\omega_n)$.
Evaluating the sum over the fermionic energy $\varepsilon$, combining two contributions together, and performing analytic continuation, we find
\begin{gather}
\Im \Pi_{s,{\rm pf}}^{R,(2),{\rm II}}(\omega)  =
 - \partial_\Delta \sum_{\sigma=\pm} \frac{f_\sigma-f_{-\sigma}}{2\omega \sigma}
Y_\sigma   \Im K_\sigma^{R,(1,1)}(\omega) \notag \\
+
  \frac{T}{8} \sum_{\sigma=\pm}  \bigl ( f^\prime_\sigma+f^\prime_{-\sigma}\bigr )
\Im K^{R,(2,2)}_\sigma(\omega)
.
 \label{eq:A:d2:1}
\end{gather}
Here we introduce the following function
\begin{equation}
Y_\sigma = T\sum_\Omega \frac{\tilde{\alpha}(i\Omega)}{i\Omega+\Delta\sigma} .
\end{equation}
Strictly speaking the sum defining $Y_\sigma$ is divergent. The summation is truncated at $\Omega=\Omega_{\rm max} \sim E_c$ which is the model cut-off. 
The function $Y_\sigma$ is therefore cut-off dependent.  It is also important to note that when evaluating summation over boson frequencies $\Omega^\prime$ and coming across divergent expressions  we assume symmetric limits $-\Omega_{\rm max}<\Omega^\prime<\Omega_{\rm max}$. 
Truncated symmetric sums allows us to repeatedly  shift a summation variable safely.

\subsection{The diagrams IIIa and IIIb}

The contribution from the diagram IIIa to the polarization operator can be  written as
\begin{gather}
\Pi_{s,{\rm pf}}^{(2),{\rm IIIa}}(i\omega_n) = - \frac{T^3 }{4} \sum_{\sigma=\pm} \sigma^2 \sum_{\varepsilon,\Omega,\Omega^\prime} \tilde{\alpha}(i\Omega)\tilde{\alpha}(i\Omega^\prime) G^3_\sigma(i\varepsilon)  \notag \\
\times
G_\sigma(i\varepsilon+i\omega_n)
G_{-\sigma}(i\varepsilon+i\Omega)
G_{-\sigma}(i\varepsilon+i\Omega^\prime)  .
 \end{gather}
The contribution from the diagram IIIb can be found from the expression above by reverting the sign of $\omega_n$: $\Pi_{s,{\rm pf}}^{(2),{\rm IIIb}}(i\omega_n)=\Pi_{s,{\rm pf}}^{(2),{\rm IIIa}}(-i\omega_n)$.
Evaluating the sum over the fermionic energy $\varepsilon$, combining two contributions together, and performing analytic continuation, we find
\begin{gather}
\Im \Pi_{s,{\rm pf}}^{R,(2),{\rm III}}(\omega)  =
\sum_{\sigma=\pm} \frac{f_\sigma-f_{-\sigma}}{2\omega}\Im \Bigl [ K_\sigma^{R,(1,1)}(\omega) \Bigr ]^2\notag \\
+
  \sum_{\sigma=\pm}  \frac{f_\sigma-f_{-\sigma}}{\omega^2}
Y_\sigma
\Im K_\sigma^{R,(1,1)}(\omega)
.
 \label{eq:A:d3:1}
\end{gather}

\subsection{The diagrams IVa, IVb, IVc, and IVd}

The contribution from the diagram IVa to the polarization operator can be  written as
\begin{gather}
\Pi_{s,{\rm pf}}^{(2),{\rm IVa}}(i\omega_n) = - \frac{T^3 }{4} \sum_{\sigma=\pm} (-\sigma^2) \sum_{\varepsilon,\Omega,\Omega^\prime} \tilde{\alpha}(i\Omega)\tilde{\alpha}(i\Omega^\prime) G^2_\sigma(i\varepsilon)  \notag \\
\times
G_\sigma(i\varepsilon+i\omega_n)
G_{-\sigma}(i\varepsilon+i\Omega^\prime) G_{-\sigma}(i\varepsilon+i\Omega) \notag \\
\times  G_{-\sigma}(i\varepsilon+i\Omega+i\omega_n).
 \end{gather}
The contribution from the other three diagrams can be found from the expression above by reverting the sign of $\omega_n$ and the sign of $\sigma$ and the summation sign: $\Pi_{s,{\rm pf}}^{(2),{\rm IVb}}(i\omega_n,\sigma)=\Pi_{s,{\rm pf}}^{(2),{\rm IVa}}(-i\omega_n,-\sigma)$,
$\Pi_{s,{\rm pf}}^{(2),{\rm IVc}}(i\omega_n,\sigma)=\Pi_{s,{\rm pf}}^{(2),{\rm IVa}}(i\omega_n,-\sigma)$,
and
$\Pi_{s,{\rm pf}}^{(2),{\rm IVd}}(i\omega_n,\sigma)=\Pi_{s,{\rm pf}}^{(2),{\rm IVa}}(-i\omega_n,\sigma)$.
Evaluating the sum over the fermionic energy $\varepsilon$, combining all four contributions together, and performing analytic continuation, we find
\begin{gather}
\Im \Pi_{s,{\rm pf}}^{R,(2),{\rm IV}}(\omega)  =
-  \partial_\Delta \sum_{\sigma=\pm} \frac{f_\sigma-f_{-\sigma}}{\omega \sigma}
Y_\sigma   \Im K_\sigma^{R,(1,1)}(\omega)
\notag\\
+ \sum_{\sigma=\pm} \frac{f_\sigma-f_{-\sigma}}{\omega}\Im \Bigl [ K_\sigma^{R,(1,1)}(\omega) \Bigr ]^2
\notag \\
  - T \sum_{\sigma=\pm} \frac{f^\prime_\sigma+f^\prime_{-\sigma}}{4\omega}
\Im \Bigl [2 K_\sigma^{R,(2,1)}(\omega) - \omega K_\sigma^{R,(2,2)}(\omega) \Bigr ]
.
 \label{eq:A:d4:1}
\end{gather}

\subsection{The diagram V}

The contribution from the diagram V to the polarization operator can be  written as
\begin{gather}
\Pi_{s,{\rm pf}}^{(2),{\rm V}}(i\omega_n) = - \frac{T^3 }{4} \sum_{\sigma=\pm} \sigma^2 \sum_{\varepsilon,\Omega,\Omega^\prime} \tilde{\alpha}(i\Omega)\tilde{\alpha}(i\Omega^\prime) G^2_\sigma(i\varepsilon)  \notag \\
\times
G^2_\sigma(i\varepsilon+i\omega_n)
G_{-\sigma}(i\varepsilon+i\Omega^\prime) G_{-\sigma}(i\varepsilon+i\Omega).
 \end{gather}
Evaluating the sum over the fermionic energy $\varepsilon$ and performing analytic continuation, we find
\begin{gather}
\Im \Pi_{s,{\rm pf}}^{R,(2),{\rm V}}(\omega)  =
- \sum_{\sigma=\pm} \frac{f_\sigma-f_{-\sigma}}{\omega^2}Y_\sigma
\Im  K_\sigma^{R,(1,1)}(\omega) \notag \\
-\partial_\Delta \sum_{\sigma=\pm} \frac{f_\sigma-f_{-\sigma}}{2\omega\sigma}Y_\sigma
\Im  K_\sigma^{R,(1,1)}(\omega)  \notag \\
  - T\sum_{\sigma=\pm} \frac{f^\prime_\sigma+f^\prime_{-\sigma}}{8\omega}
\Im \Bigl [2 K_\sigma^{R,(2,1)}(\omega) - \omega K_\sigma^{R,(2,2)}(\omega) \Bigr ]
.
 \label{eq:A:d5:1}
\end{gather}

\begin{table}[t]
\caption{The first few integrals $I_k$.}
\begin{tabular}{l}
$I_1 =  - \omega(\omega^2 +4\pi^2 T^2)/3$ \\
$I_2 = \omega^2(\omega^2 +4\pi^2 T^2)/6$ \\
$I_3 = -\omega (3\omega^4 + 20\pi^2 T^2 \omega^2 + 32\pi^4 T^4)/30$ \\
$I_4 = \omega^2 ( \omega^4 + 10\pi^2 T^2 \omega^2 + 24 \pi^4 T^4)/15$
\end{tabular}
\label{Tab}
\end{table}

\subsection{The result for $\Im \Pi_{s,{\rm pf}}^{R}(\omega)$ in the 2d order in $g$}

Combining all contributions, Eqs. \eqref{eq:A:d1:1}, \eqref{eq:A:d2:1}, \eqref{eq:A:d3:1}, \eqref{eq:A:d4:1}, and \eqref{eq:A:d5:1}, together, we find
\begin{gather}
 \Im \Pi_{s,{\rm pf}}^{R,(2)}(\omega) =
\frac{T}{2}\sum_{\sigma} \bigl ( f^\prime_\sigma+f^\prime_{-\sigma} \bigr )  \Im K_\sigma^{R,(2,2)}(\omega) \notag \\
-2 \partial_\Delta \sum_{\sigma=\pm} \frac{f_\sigma-f_{-\sigma}}{\omega\sigma}Y_\sigma
\Im  K_\sigma^{R,(1,1)}(\omega) \notag \\
 - T\sum_{\sigma} \frac{f^\prime_\sigma+f^\prime_{-\sigma}}{\omega}  \Im K_\sigma^{R,(2,1)}(\omega)
 \notag \\
+  2 \sum_{\sigma=\pm} \frac{f_\sigma-f_{-\sigma}}{\omega}\Im \Bigl [ K_\sigma^{R,(1,1)}(\omega) \Bigr ]^2
   .
  \label{eq:A:f:1a}
\end{gather}
The following comment is in order here. Alternatively, one can compute the admittance by means of the current-current correlation function [\onlinecite{Nazarov1990}]. The latter consists of two operators which are of the first and second order in $g$ [\onlinecite{Ben-Jacob1983}]. Then the first and second lines in Eq. \eqref{eq:A:f:1}  comes from the renormalization of the operator of the first order in $g$ whereas the third and forth lines correspond to the contributions from the operator of the second order in $g$. Using Eq. \eqref{eq:zero:order}, we find
\begin{gather}
 \Im \Pi_{s}^{R,(2)}(\omega) =
-2 \tanh\frac{\beta\Delta}{2} \sum_{\sigma=\pm} \frac{\sigma}{\omega}\Im \Bigl [ K_\sigma^{R,(1,1)}(\omega) \Bigr ]^2
 \notag \\
 +\frac{2}{\cosh(\beta\Delta/2)} \partial_\Delta \sum_{\sigma=\pm} \frac{\sinh(\beta\Delta/2)}{\omega}Y_\sigma
\Im  K_\sigma^{R,(1,1)}(\omega)  \notag \\
 + \sum_{\sigma}  \Im \Bigl [\frac{1}{2} K_\sigma^{R,(2,2)} - \frac{1}{\omega} K_\sigma^{R,(2,1)}(\omega)  \Bigr ]
  .
  \label{eq:A:f:1}
\end{gather}
This expression is convenient for analysis at $|\omega|, T \ll |\Delta|$.  Converting the sum over Matsubara frequencies in Eq. \eqref{eq:A:def:K}  into the integral and performing analytic continuation, $i\omega \to \omega+i0$, we obtain
\begin{gather}
\Im K_\sigma^{R,(1,1)}(\omega) =
\int\limits_{-\infty}^\infty \frac{d \varepsilon}{2\pi\omega}
\bigl ( \mathcal{B}_\varepsilon - \mathcal{B}_{\varepsilon+\omega}\bigr )
\Im D_\sigma^R(\varepsilon) \notag \\
\times \Im \tilde{\alpha}^R(\varepsilon+\omega)  .
\end{gather}
Here $\mathcal{B}_\varepsilon= \coth (\varepsilon/2T)$, $\tilde{\alpha}^R(\omega) = - \tilde{\alpha}^A(\omega)  =- i g \omega/(4\pi)$, and $D_\sigma^R(\varepsilon) = [\varepsilon+\Delta\sigma+i0]^{-1}$.
Since $\Im D_\sigma^R(\varepsilon) = -\pi \delta(\varepsilon+\Delta\sigma)$, the imaginary part of $K_\sigma^{R,(1,1)}$ is exponentially small, $\sim \exp(-|\Delta|/T)$.
We note that $K_\sigma^{R,(1,1)}$ is defined by the divergent Matsubara sum. It should be understood as a finite sum truncated at $\Omega=\Omega_{\rm max}\sim E_c$.
Then the analytical continuation is possible. Moreover the imaginary part of $K_\sigma^{R,(1,1)}$ is $E_c$ independent.
 
 The terms proportional to $K_\sigma^{R,(1,1)}$ are responsible for the renormalization of the first order perturbative result \eqref{eq:1o:g}. Therefore, only the last line in Eq. \eqref{eq:A:f:1} contributes to $\Im \Pi_{s}^{R,(2)}(\omega)$ in the regime $|\omega|, T \ll |\Delta|$.

 Again converting the sun over Matsubara frequencies in Eq. \eqref{eq:A:def:K}  into the integral and performing analytic continuation, $i\omega \to \omega+i0$, we find
\begin{gather}
\Im K_\sigma^{R,(2,1)}(\omega)   =  \int\limits_{-\infty}^\infty \frac{d \varepsilon}{\pi
\omega^2} \bigl ( \mathcal{B}_{\varepsilon+\omega}-\mathcal{B}_\varepsilon \bigr ) \Re D_\sigma^R (\varepsilon) \Im \tilde{\alpha}^R(\varepsilon) \notag \\
\times \Im \tilde{\alpha}^R(\varepsilon+\omega)
=
\left (\frac{g}{4\pi} \right )^2 \sum_{k=1}^\infty \frac{(-1)^{k-1} I_k}{\pi\omega^2 (\Delta \sigma)^k} .
\end{gather}
Here we perform expansion in series in $1/\Delta$. The functions $I_k$ are defined as follows
\begin{equation}
I_k = \int \limits_{-\infty}^\infty d\varepsilon \, \varepsilon^k(\varepsilon+\omega)\bigl (\mathcal{B}_{\varepsilon+\omega}-\mathcal{B}_\varepsilon\bigr ) .
\end{equation}
We list several first functions $I_k$ (for $k=1,2,3,4$) in Table \ref{Tab}. We note that the very same functions $I_k$ determine the imaginary part of $K_\sigma^{R,(2,2)}(\omega)$:
\begin{gather}
\Im K_\sigma^{R,(2,2)}(\omega)  = - \sigma \partial_\Delta \Im K_\sigma^{R,(2,2)}(\omega)
\notag \\
= \left (\frac{g}{4\pi} \right )^2 \sum_{k=1}^\infty \frac{ k (-1)^{k-1} I_k}{\pi\omega^2 (\Delta \sigma)^{k+1}} .
\end{gather}
Using the functions $I_k$ from the Table \ref{Tab}, from Eq. \eqref{eq:A:f:1} we obtain the result \eqref{eq:PRR:2} in the main text.

Finally, we note that for the case of $q=k$, i.e. at the Coulomb peaks, one cannot adopt the pseudofermion technique and has to take into account all charging states. Then, the contribution to $\Re \mathcal{G}(\omega)$ due to the inelastic cotunneling is given by the last line of Eq. \eqref{eq:A:f:1} multiplied by a factor of 2 and with the following substitutions: $\Delta \to E_c$ and
\begin{equation}
\frac{1}{\omega} \to \frac{1}{\omega} \left (1+\frac{\omega^2}{2E_c(2E_c-\omega\sigma)} \right ).
\end{equation}
It is due to this additional term the contribution of the order of $1/E_c^4$ cancels and the admittance becomes proportional to $1/E_c^6$ [\onlinecite{Nazarov1990}].

\bibliography{paper-biblio}

\begin{thebibliography}{50}%
\makeatletter
\providecommand \@ifxundefined [1]{%
 \@ifx{#1\undefined}
}%
\providecommand \@ifnum [1]{%
 \ifnum #1\expandafter \@firstoftwo
 \else \expandafter \@secondoftwo
 \fi
}%
\providecommand \@ifx [1]{%
 \ifx #1\expandafter \@firstoftwo
 \else \expandafter \@secondoftwo
 \fi
}%
\providecommand \natexlab [1]{#1}%
\providecommand \enquote  [1]{``#1''}%
\providecommand \bibnamefont  [1]{#1}%
\providecommand \bibfnamefont [1]{#1}%
\providecommand \citenamefont [1]{#1}%
\providecommand \href@noop [0]{\@secondoftwo}%
\providecommand \href [0]{\begingroup \@sanitize@url \@href}%
\providecommand \@href[1]{\@@startlink{#1}\@@href}%
\providecommand \@@href[1]{\endgroup#1\@@endlink}%
\providecommand \@sanitize@url [0]{\catcode `\\12\catcode `\$12\catcode
  `\&12\catcode `\#12\catcode `\^12\catcode `\_12\catcode `\%12\relax}%
\providecommand \@@startlink[1]{}%
\providecommand \@@endlink[0]{}%
\providecommand \url  [0]{\begingroup\@sanitize@url \@url }%
\providecommand \@url [1]{\endgroup\@href {#1}{\urlprefix }}%
\providecommand \urlprefix  [0]{URL }%
\providecommand \Eprint [0]{\href }%
\providecommand \doibase [0]{http://dx.doi.org/}%
\providecommand \selectlanguage [0]{\@gobble}%
\providecommand \bibinfo  [0]{\@secondoftwo}%
\providecommand \bibfield  [0]{\@secondoftwo}%
\providecommand \translation [1]{[#1]}%
\providecommand \BibitemOpen [0]{}%
\providecommand \bibitemStop [0]{}%
\providecommand \bibitemNoStop [0]{.\EOS\space}%
\providecommand \EOS [0]{\spacefactor3000\relax}%
\providecommand \BibitemShut  [1]{\csname bibitem#1\endcsname}%
\let\auto@bib@innerbib\@empty
\bibitem [{\citenamefont {Sch\"{o}n}\ and\ \citenamefont
  {Zaikin}(1990)}]{zaikin}%
  \BibitemOpen
  \bibfield  {author} {\bibinfo {author} {\bibfnamefont {G.}~\bibnamefont
  {Sch\"{o}n}}\ and\ \bibinfo {author} {\bibfnamefont {A.}~\bibnamefont
  {Zaikin}},\ }\bibfield  {title} {\enquote {\bibinfo {title} {Quantum coherent
  effects, phase transitions, and the dissipative dynamics of ultra small
  tunnel junctions},}\ }\href
  {http://www.sciencedirect.com/science/article/pii/037015739090156V}
  {\bibfield  {journal} {\bibinfo  {journal} {Phys. Rep.}\ }\textbf {\bibinfo
  {volume} {198}},\ \bibinfo {pages} {237} (\bibinfo {year}
  {1990})}\BibitemShut {NoStop}%
\bibitem [{ZPh(1991, special issue on single charge tunneling, ed. by H.
  Grabert and H. Horner.)}]{ZPhys}%
  \BibitemOpen
  \href@noop {} {\bibfield  {journal} {\bibinfo  {journal} {Z. Phys. B:
  Condens. Matter}\ }\textbf {\bibinfo {volume} {85}},\ \bibinfo {pages} {317}
  (\bibinfo {year} {1991, special issue on single charge tunneling, ed. by H.
  Grabert and H. Horner.})}\BibitemShut {NoStop}%
\bibitem [{\citenamefont {Grabert}\ and\ \citenamefont
  {Devoret}(1992)}]{grabert}%
  \BibitemOpen
  \bibinfo {editor} {\bibfnamefont {H.}~\bibnamefont {Grabert}}\ and\ \bibinfo
  {editor} {\bibfnamefont {M.~H.}\ \bibnamefont {Devoret}},\ eds.,\ \href@noop
  {} {\emph {\bibinfo {title} {Single Charge Tunneling}}}\ (\bibinfo
  {publisher} {Plenum},\ \bibinfo {address} {New York},\ \bibinfo {year}
  {1992})\BibitemShut {NoStop}%
\bibitem [{\citenamefont {Blanter}\ and\ \citenamefont
  {B\"{u}ttiker}(2000)}]{blanter}%
  \BibitemOpen
  \bibfield  {author} {\bibinfo {author} {\bibfnamefont {Y.}~\bibnamefont
  {Blanter}}\ and\ \bibinfo {author} {\bibfnamefont {M.}~\bibnamefont
  {B\"{u}ttiker}},\ }\bibfield  {title} {\enquote {\bibinfo {title} {Shot noise
  in mesoscopic conductors},}\ }\href
  {http://www.sciencedirect.com/science/article/pii/S0370157399001234}
  {\bibfield  {journal} {\bibinfo  {journal} {Phys. Rep.}\ }\textbf {\bibinfo
  {volume} {336}},\ \bibinfo {pages} {1} (\bibinfo {year} {2000})}\BibitemShut
  {NoStop}%
\bibitem [{\citenamefont {Aleiner}\ \emph {et~al.}(2002)\citenamefont
  {Aleiner}, \citenamefont {Brouwer},\ and\ \citenamefont {Glazman}}]{aleiner}%
  \BibitemOpen
  \bibfield  {author} {\bibinfo {author} {\bibfnamefont {I.}~\bibnamefont
  {Aleiner}}, \bibinfo {author} {\bibfnamefont {P.}~\bibnamefont {Brouwer}}, \
  and\ \bibinfo {author} {\bibfnamefont {L.}~\bibnamefont {Glazman}},\
  }\bibfield  {title} {\enquote {\bibinfo {title} {{Quantum effects in Coulomb
  blockade}},}\ }\href
  {http://www.sciencedirect.com/science/article/pii/S0370157301000631}
  {\bibfield  {journal} {\bibinfo  {journal} {Phys. Rep.}\ }\textbf {\bibinfo
  {volume} {358}},\ \bibinfo {pages} {309} (\bibinfo {year}
  {2002})}\BibitemShut {NoStop}%
\bibitem [{\citenamefont {Glazman}\ and\ \citenamefont
  {Pustilnik}(2003)}]{Glazman}%
  \BibitemOpen
  \bibfield  {author} {\bibinfo {author} {\bibfnamefont {L.~I.}\ \bibnamefont
  {Glazman}}\ and\ \bibinfo {author} {\bibfnamefont {M.}~\bibnamefont
  {Pustilnik}},\ }\bibfield  {title} {\enquote {\bibinfo {title}
  {Low-temperature transport through a quantum dot},}\ }in\ \href@noop {}
  {\emph {\bibinfo {booktitle} {New Directions in Mesoscopic Physics (Towards
  to Nanoscience)}}},\ \bibinfo {editor} {edited by\ \bibinfo {editor}
  {\bibfnamefont {R.}~\bibnamefont {Fazio}}, \bibinfo {editor} {\bibfnamefont
  {V.~F.}\ \bibnamefont {Gantmakher}}, \ and\ \bibinfo {editor} {\bibfnamefont
  {Y.}~\bibnamefont {Imry}}}\ (\bibinfo  {publisher} {Kluwer},\ \bibinfo
  {address} {Dordrecht},\ \bibinfo {year} {2003})\BibitemShut {NoStop}%
\bibitem [{\citenamefont {Nazarov}(1990)}]{Nazarov1990}%
  \BibitemOpen
  \bibfield  {author} {\bibinfo {author} {\bibfnamefont {Yu.~V.}\ \bibnamefont
  {Nazarov}},\ }\bibfield  {title} {\enquote {\bibinfo {title} {Dissipation in
  a {Coulomb} capacitor},}\ }\href@noop {} {\bibfield  {journal} {\bibinfo
  {journal} {Sov. J. Low Temp. Phys.}\ }\textbf {\bibinfo {volume} {16}},\
  \bibinfo {pages} {422} (\bibinfo {year} {1990})}\BibitemShut {NoStop}%
\bibitem [{\citenamefont {Matveev}(1991)}]{matveev}%
  \BibitemOpen
  \bibfield  {author} {\bibinfo {author} {\bibfnamefont {K.~A.}\ \bibnamefont
  {Matveev}},\ }\bibfield  {title} {\enquote {\bibinfo {title} {{Quantum
  fluctuations of the charge of a metal particle under the Coulomb blockade
  conditions}},}\ }\href@noop {} {\bibfield  {journal} {\bibinfo  {journal}
  {Sov. Phys. JETP}\ }\textbf {\bibinfo {volume} {72}},\ \bibinfo {pages} {892}
  (\bibinfo {year} {1991})}\BibitemShut {NoStop}%
\bibitem [{\citenamefont {Grabert}(1994{\natexlab{a}})}]{Grabert0a}%
  \BibitemOpen
  \bibfield  {author} {\bibinfo {author} {\bibfnamefont {H.}~\bibnamefont
  {Grabert}},\ }\bibfield  {title} {\enquote {\bibinfo {title} {{Rounding of
  the Coulomb staircase by the tunneling conductance}},}\ }\href@noop {}
  {\bibfield  {journal} {\bibinfo  {journal} {Physica B}\ }\textbf {\bibinfo
  {volume} {194}},\ \bibinfo {pages} {1011} (\bibinfo {year}
  {1994}{\natexlab{a}})}\BibitemShut {NoStop}%
\bibitem [{\citenamefont {Grabert}(1994{\natexlab{b}})}]{Grabert0b}%
  \BibitemOpen
  \bibfield  {author} {\bibinfo {author} {\bibfnamefont {H.}~\bibnamefont
  {Grabert}},\ }\bibfield  {title} {\enquote {\bibinfo {title} {{Charge
  fluctuations in the single-electron box: Perturbation expansion in the
  tunneling conductance}},}\ }\href {\doibase 10.1103/PhysRevB.50.17364}
  {\bibfield  {journal} {\bibinfo  {journal} {Phys. Rev. B}\ }\textbf {\bibinfo
  {volume} {50}},\ \bibinfo {pages} {17364} (\bibinfo {year}
  {1994}{\natexlab{b}})}\BibitemShut {NoStop}%
\bibitem [{\citenamefont {Matveev}(1995)}]{matveev1}%
  \BibitemOpen
  \bibfield  {author} {\bibinfo {author} {\bibfnamefont {K.~A.}\ \bibnamefont
  {Matveev}},\ }\bibfield  {title} {\enquote {\bibinfo {title} {Coulomb
  blockade at almost perfect transmission},}\ }\href {\doibase
  10.1103/PhysRevB.51.1743} {\bibfield  {journal} {\bibinfo  {journal} {Phys.
  Rev. B}\ }\textbf {\bibinfo {volume} {51}},\ \bibinfo {pages} {1743}
  (\bibinfo {year} {1995})}\BibitemShut {NoStop}%
\bibitem [{\citenamefont {Wang}\ and\ \citenamefont
  {Grabert}(1996)}]{Grabert1}%
  \BibitemOpen
  \bibfield  {author} {\bibinfo {author} {\bibfnamefont {X.}~\bibnamefont
  {Wang}}\ and\ \bibinfo {author} {\bibfnamefont {H.}~\bibnamefont {Grabert}},\
  }\bibfield  {title} {\enquote {\bibinfo {title} {Coulomb charging at large
  conduction},}\ }\href {\doibase 10.1103/PhysRevB.53.12621} {\bibfield
  {journal} {\bibinfo  {journal} {Phys. Rev. B}\ }\textbf {\bibinfo {volume}
  {53}},\ \bibinfo {pages} {12621} (\bibinfo {year} {1996})}\BibitemShut
  {NoStop}%
\bibitem [{\citenamefont {G\"oppert}\ \emph {et~al.}(1998)\citenamefont
  {G\"oppert}, \citenamefont {Grabert}, \citenamefont {Prokof'ev},\ and\
  \citenamefont {Svistunov}}]{Grabert2}%
  \BibitemOpen
  \bibfield  {author} {\bibinfo {author} {\bibfnamefont {G.}~\bibnamefont
  {G\"oppert}}, \bibinfo {author} {\bibfnamefont {H.}~\bibnamefont {Grabert}},
  \bibinfo {author} {\bibfnamefont {N.~V.}\ \bibnamefont {Prokof'ev}}, \ and\
  \bibinfo {author} {\bibfnamefont {B.~V.}\ \bibnamefont {Svistunov}},\
  }\bibfield  {title} {\enquote {\bibinfo {title} {{Effect of tunneling
  conductance on the Coulomb staircase}},}\ }\href {\doibase
  10.1103/PhysRevLett.81.2324} {\bibfield  {journal} {\bibinfo  {journal}
  {Phys. Rev. Lett.}\ }\textbf {\bibinfo {volume} {81}},\ \bibinfo {pages}
  {2324} (\bibinfo {year} {1998})}\BibitemShut {NoStop}%
\bibitem [{\citenamefont {Beloborodov}\ \emph {et~al.}(2003)\citenamefont
  {Beloborodov}, \citenamefont {Andreev},\ and\ \citenamefont
  {Larkin}}]{beloborodov1}%
  \BibitemOpen
  \bibfield  {author} {\bibinfo {author} {\bibfnamefont {I.~S.}\ \bibnamefont
  {Beloborodov}}, \bibinfo {author} {\bibfnamefont {A.~V.}\ \bibnamefont
  {Andreev}}, \ and\ \bibinfo {author} {\bibfnamefont {A.~I.}\ \bibnamefont
  {Larkin}},\ }\bibfield  {title} {\enquote {\bibinfo {title} {{Two-loop
  approximation in the Coulomb blockade problem}},}\ }\href {\doibase
  10.1103/PhysRevB.68.024204} {\bibfield  {journal} {\bibinfo  {journal} {Phys.
  Rev. B}\ }\textbf {\bibinfo {volume} {68}},\ \bibinfo {pages} {024204}
  (\bibinfo {year} {2003})}\BibitemShut {NoStop}%
\bibitem [{\citenamefont {B\"{u}ttiker}\ \emph {et~al.}(1993)\citenamefont
  {B\"{u}ttiker}, \citenamefont {Thomas},\ and\ \citenamefont
  {Pretre}}]{buttiker0}%
  \BibitemOpen
  \bibfield  {author} {\bibinfo {author} {\bibfnamefont {M.}~\bibnamefont
  {B\"{u}ttiker}}, \bibinfo {author} {\bibfnamefont {H.}~\bibnamefont
  {Thomas}}, \ and\ \bibinfo {author} {\bibfnamefont {A.}~\bibnamefont
  {Pretre}},\ }\bibfield  {title} {\enquote {\bibinfo {title} {Mesoscopic
  capacitors},}\ }\href
  {http://www.sciencedirect.com/science/article/pii/0375960193911939}
  {\bibfield  {journal} {\bibinfo  {journal} {Phys. Lett. A}\ }\textbf
  {\bibinfo {volume} {180}},\ \bibinfo {pages} {364} (\bibinfo {year}
  {1993})}\BibitemShut {NoStop}%
\bibitem [{\citenamefont {B\"uttiker}\ and\ \citenamefont
  {Martin}(2000)}]{buttiker3}%
  \BibitemOpen
  \bibfield  {author} {\bibinfo {author} {\bibfnamefont {M.}~\bibnamefont
  {B\"uttiker}}\ and\ \bibinfo {author} {\bibfnamefont {A.~M.}\ \bibnamefont
  {Martin}},\ }\bibfield  {title} {\enquote {\bibinfo {title} {{Charge
  relaxation and dephasing in Coulomb-coupled conductors}},}\ }\href {\doibase
  10.1103/PhysRevB.61.2737} {\bibfield  {journal} {\bibinfo  {journal} {Phys.
  Rev. B}\ }\textbf {\bibinfo {volume} {61}},\ \bibinfo {pages} {2737}
  (\bibinfo {year} {2000})}\BibitemShut {NoStop}%
\bibitem [{\citenamefont {Nigg}\ \emph {et~al.}(2006)\citenamefont {Nigg},
  \citenamefont {L\'opez},\ and\ \citenamefont {B\"uttiker}}]{buttiker2}%
  \BibitemOpen
  \bibfield  {author} {\bibinfo {author} {\bibfnamefont {S.~E.}\ \bibnamefont
  {Nigg}}, \bibinfo {author} {\bibfnamefont {R.}~\bibnamefont {L\'opez}}, \
  and\ \bibinfo {author} {\bibfnamefont {M.}~\bibnamefont {B\"uttiker}},\
  }\bibfield  {title} {\enquote {\bibinfo {title} {Mesoscopic charge
  relaxation},}\ }\href {\doibase 10.1103/PhysRevLett.97.206804} {\bibfield
  {journal} {\bibinfo  {journal} {Phys. Rev. Lett.}\ }\textbf {\bibinfo
  {volume} {97}},\ \bibinfo {pages} {206804} (\bibinfo {year}
  {2006})}\BibitemShut {NoStop}%
\bibitem [{\citenamefont {Nigg}\ and\ \citenamefont
  {B\"uttiker}(2008)}]{buttiker1}%
  \BibitemOpen
  \bibfield  {author} {\bibinfo {author} {\bibfnamefont {S.~E.}\ \bibnamefont
  {Nigg}}\ and\ \bibinfo {author} {\bibfnamefont {M.}~\bibnamefont
  {B\"uttiker}},\ }\bibfield  {title} {\enquote {\bibinfo {title} {Quantum to
  classical transition of the charge relaxation resistance of a mesoscopic
  capacitor},}\ }\href {\doibase 10.1103/PhysRevB.77.085312} {\bibfield
  {journal} {\bibinfo  {journal} {Phys. Rev. B}\ }\textbf {\bibinfo {volume}
  {77}},\ \bibinfo {pages} {085312} (\bibinfo {year} {2008})}\BibitemShut
  {NoStop}%
\bibitem [{\citenamefont {Ringel}\ \emph {et~al.}(2008)\citenamefont {Ringel},
  \citenamefont {Imry},\ and\ \citenamefont {Entin-Wohlman}}]{imry}%
  \BibitemOpen
  \bibfield  {author} {\bibinfo {author} {\bibfnamefont {Z.}~\bibnamefont
  {Ringel}}, \bibinfo {author} {\bibfnamefont {Y.}~\bibnamefont {Imry}}, \ and\
  \bibinfo {author} {\bibfnamefont {O.}~\bibnamefont {Entin-Wohlman}},\
  }\bibfield  {title} {\enquote {\bibinfo {title} {Delayed currents and
  interaction effects in mesoscopic capacitors},}\ }\href {\doibase
  10.1103/PhysRevB.78.165304} {\bibfield  {journal} {\bibinfo  {journal} {Phys.
  Rev. B}\ }\textbf {\bibinfo {volume} {78}},\ \bibinfo {pages} {165304}
  (\bibinfo {year} {2008})}\BibitemShut {NoStop}%
\bibitem [{\citenamefont {Park}\ and\ \citenamefont {Ahn}(2008)}]{Park}%
  \BibitemOpen
  \bibfield  {author} {\bibinfo {author} {\bibfnamefont {H.~C.}\ \bibnamefont
  {Park}}\ and\ \bibinfo {author} {\bibfnamefont {K.-H.}\ \bibnamefont {Ahn}},\
  }\bibfield  {title} {\enquote {\bibinfo {title} {{Admittance and noise in an
  electrically driven nanostructure: Interplay between quantum coherence and
  statistics}},}\ }\href {\doibase 10.1103/PhysRevLett.101.116804} {\bibfield
  {journal} {\bibinfo  {journal} {Phys. Rev. Lett.}\ }\textbf {\bibinfo
  {volume} {101}},\ \bibinfo {pages} {116804} (\bibinfo {year}
  {2008})}\BibitemShut {NoStop}%
\bibitem [{\citenamefont {Gabelli}\ \emph {et~al.}(2006)\citenamefont
  {Gabelli}, \citenamefont {Feve}, \citenamefont {Berroir}, \citenamefont
  {Pla\c{c}ais}, \citenamefont {Cavanna}, \citenamefont {Etienne},
  \citenamefont {Jin},\ and\ \citenamefont {Glattli}}]{gabelli}%
  \BibitemOpen
  \bibfield  {author} {\bibinfo {author} {\bibfnamefont {J.}~\bibnamefont
  {Gabelli}}, \bibinfo {author} {\bibfnamefont {G.}~\bibnamefont {Feve}},
  \bibinfo {author} {\bibfnamefont {J.-M.}\ \bibnamefont {Berroir}}, \bibinfo
  {author} {\bibfnamefont {B.}~\bibnamefont {Pla\c{c}ais}}, \bibinfo {author}
  {\bibfnamefont {A.}~\bibnamefont {Cavanna}}, \bibinfo {author} {\bibfnamefont
  {B.}~\bibnamefont {Etienne}}, \bibinfo {author} {\bibfnamefont
  {Y.}~\bibnamefont {Jin}}, \ and\ \bibinfo {author} {\bibfnamefont {D.~C.}\
  \bibnamefont {Glattli}},\ }\bibfield  {title} {\enquote {\bibinfo {title}
  {{Violation of Kirchhoff's laws for a coherent $\ensuremath{RC}$ circuit}},}\
  }\href {\doibase 10.1126/science.1126940} {\bibfield  {journal} {\bibinfo
  {journal} {Science}\ }\textbf {\bibinfo {volume} {313}},\ \bibinfo {pages}
  {499} (\bibinfo {year} {2006})}\BibitemShut {NoStop}%
\bibitem [{\citenamefont {Persson}\ \emph {et~al.}(2010)\citenamefont
  {Persson}, \citenamefont {Wilson}, \citenamefont {Sandberg}, \citenamefont
  {Johansson},\ and\ \citenamefont {Delsing}}]{delsing}%
  \BibitemOpen
  \bibfield  {author} {\bibinfo {author} {\bibfnamefont {F.}~\bibnamefont
  {Persson}}, \bibinfo {author} {\bibfnamefont {C.~M.}\ \bibnamefont {Wilson}},
  \bibinfo {author} {\bibfnamefont {M.}~\bibnamefont {Sandberg}}, \bibinfo
  {author} {\bibfnamefont {G.}~\bibnamefont {Johansson}}, \ and\ \bibinfo
  {author} {\bibfnamefont {P.}~\bibnamefont {Delsing}},\ }\bibfield  {title}
  {\enquote {\bibinfo {title} {{Excess dissipation in a dingle-electron box:
  The Sisyphus resistance}},}\ }\href {\doibase 10.1021/nl903887x} {\bibfield
  {journal} {\bibinfo  {journal} {Nano Lett.}\ }\textbf {\bibinfo {volume}
  {10}},\ \bibinfo {pages} {953--957} (\bibinfo {year} {2010})}\BibitemShut
  {NoStop}%
\bibitem [{\citenamefont {Ciccarelli}\ and\ \citenamefont
  {Ferguson}(2011)}]{Ciccarelli}%
  \BibitemOpen
  \bibfield  {author} {\bibinfo {author} {\bibfnamefont {C.}~\bibnamefont
  {Ciccarelli}}\ and\ \bibinfo {author} {\bibfnamefont {A.~J.}\ \bibnamefont
  {Ferguson}},\ }\bibfield  {title} {\enquote {\bibinfo {title} {Impedance of
  the single-electron transistor at radio-frequencies},}\ }\href
  {http://stacks.iop.org/1367-2630/13/i=9/a=093015} {\bibfield  {journal}
  {\bibinfo  {journal} {New J. Phys.}\ }\textbf {\bibinfo {volume} {13}},\
  \bibinfo {pages} {093015} (\bibinfo {year} {2011})}\BibitemShut {NoStop}%
\bibitem [{\citenamefont {Chorley}\ \emph {et~al.}(2012)\citenamefont
  {Chorley}, \citenamefont {Wabnig}, \citenamefont {Penfold-Fitch},
  \citenamefont {Petersson}, \citenamefont {Frake}, \citenamefont {Smith},\
  and\ \citenamefont {Buitelaar}}]{Chorley}%
  \BibitemOpen
  \bibfield  {author} {\bibinfo {author} {\bibfnamefont {S.~J.}\ \bibnamefont
  {Chorley}}, \bibinfo {author} {\bibfnamefont {J.}~\bibnamefont {Wabnig}},
  \bibinfo {author} {\bibfnamefont {Z.~V.}\ \bibnamefont {Penfold-Fitch}},
  \bibinfo {author} {\bibfnamefont {K.~D.}\ \bibnamefont {Petersson}}, \bibinfo
  {author} {\bibfnamefont {J.}~\bibnamefont {Frake}}, \bibinfo {author}
  {\bibfnamefont {C.~G.}\ \bibnamefont {Smith}}, \ and\ \bibinfo {author}
  {\bibfnamefont {M.~R.}\ \bibnamefont {Buitelaar}},\ }\bibfield  {title}
  {\enquote {\bibinfo {title} {Measuring the complex admittance of a carbon
  nanotube double quantum dot},}\ }\href {\doibase
  10.1103/PhysRevLett.108.036802} {\bibfield  {journal} {\bibinfo  {journal}
  {Phys. Rev. Lett.}\ }\textbf {\bibinfo {volume} {108}},\ \bibinfo {pages}
  {036802} (\bibinfo {year} {2012})}\BibitemShut {NoStop}%
\bibitem [{\citenamefont {Frake}\ \emph {et~al.}(2015)\citenamefont {Frake},
  \citenamefont {Kano}, \citenamefont {Ciccarelli}, \citenamefont {Griffiths},
  \citenamefont {Sakamoto}, \citenamefont {Teranishi}, \citenamefont {Majima},
  \citenamefont {Smith},\ and\ \citenamefont {Buitelaar}}]{Frake}%
  \BibitemOpen
  \bibfield  {author} {\bibinfo {author} {\bibfnamefont {J.~C.}\ \bibnamefont
  {Frake}}, \bibinfo {author} {\bibfnamefont {S.}~\bibnamefont {Kano}},
  \bibinfo {author} {\bibfnamefont {C.}~\bibnamefont {Ciccarelli}}, \bibinfo
  {author} {\bibfnamefont {J.}~\bibnamefont {Griffiths}}, \bibinfo {author}
  {\bibfnamefont {M.}~\bibnamefont {Sakamoto}}, \bibinfo {author}
  {\bibfnamefont {T.}~\bibnamefont {Teranishi}}, \bibinfo {author}
  {\bibfnamefont {Y.}~\bibnamefont {Majima}}, \bibinfo {author} {\bibfnamefont
  {C.~G.}\ \bibnamefont {Smith}}, \ and\ \bibinfo {author} {\bibfnamefont
  {M.~R.}\ \bibnamefont {Buitelaar}},\ }\bibfield  {title} {\enquote {\bibinfo
  {title} {{Radio-frequency capacitance spectroscopy of metallic
  nanoparticles}},}\ }\href
  {http://www.nature.com/srep/2015/150603/srep10858/full/srep10858.html}
  {\bibfield  {journal} {\bibinfo  {journal} {Sci. Rep.}\ }\textbf {\bibinfo
  {volume} {5}},\ \bibinfo {pages} {10858} (\bibinfo {year}
  {2015})}\BibitemShut {NoStop}%
\bibitem [{\citenamefont {Mora}\ and\ \citenamefont {Le~Hur}(2010)}]{Mora2010}%
  \BibitemOpen
  \bibfield  {author} {\bibinfo {author} {\bibfnamefont {C.}~\bibnamefont
  {Mora}}\ and\ \bibinfo {author} {\bibfnamefont {K.}~\bibnamefont {Le~Hur}},\
  }\bibfield  {title} {\enquote {\bibinfo {title} {Universal resistances of the
  quantum resistance-capacitance circuit},}\ }\href@noop {} {\bibfield
  {journal} {\bibinfo  {journal} {Nat. Phys.}\ }\textbf {\bibinfo {volume}
  {6}},\ \bibinfo {pages} {697} (\bibinfo {year} {2010})}\BibitemShut {NoStop}%
\bibitem [{\citenamefont {Korringa}(1950)}]{Korringa1950}%
  \BibitemOpen
  \bibfield  {author} {\bibinfo {author} {\bibfnamefont {J.}~\bibnamefont
  {Korringa}},\ }\bibfield  {title} {\enquote {\bibinfo {title} {Nuclear
  magnetic relaxation and resonance line shift in metals},}\ }\href@noop {}
  {\bibfield  {journal} {\bibinfo  {journal} {Physica}\ }\textbf {\bibinfo
  {volume} {16}},\ \bibinfo {pages} {601} (\bibinfo {year} {1950})}\BibitemShut
  {NoStop}%
\bibitem [{\citenamefont {Shiba}(1975)}]{Shiba1975}%
  \BibitemOpen
  \bibfield  {author} {\bibinfo {author} {\bibfnamefont {H.}~\bibnamefont
  {Shiba}},\ }\bibfield  {title} {\enquote {\bibinfo {title} {The {Korringa}
  relation for the impurity nuclear spin-lattice relaxation in dilute {Kondo}
  alloys},}\ }\href@noop {} {\bibfield  {journal} {\bibinfo  {journal} {Prog.
  Theor. Phys.}\ }\textbf {\bibinfo {volume} {54}},\ \bibinfo {pages} {967}
  (\bibinfo {year} {1975})}\BibitemShut {NoStop}%
\bibitem [{\citenamefont {Filippone}\ and\ \citenamefont
  {Mora}(2012)}]{Filippone2012}%
  \BibitemOpen
  \bibfield  {author} {\bibinfo {author} {\bibfnamefont {M.}~\bibnamefont
  {Filippone}}\ and\ \bibinfo {author} {\bibfnamefont {C.}~\bibnamefont
  {Mora}},\ }\bibfield  {title} {\enquote {\bibinfo {title} {{Fermi liquid
  approach to the quantum $\ensuremath{RC}$ circuit: Renormalization group
  analysis of the Anderson and Coulomb blockade models}},}\ }\href {\doibase
  10.1103/PhysRevB.86.125311} {\bibfield  {journal} {\bibinfo  {journal} {Phys.
  Rev. B}\ }\textbf {\bibinfo {volume} {86}},\ \bibinfo {pages} {125311}
  (\bibinfo {year} {2012})}\BibitemShut {NoStop}%
\bibitem [{\citenamefont {Dutt}\ \emph {et~al.}(2013)\citenamefont {Dutt},
  \citenamefont {Schmidt}, \citenamefont {Mora},\ and\ \citenamefont
  {Le~Hur}}]{Mora2013}%
  \BibitemOpen
  \bibfield  {author} {\bibinfo {author} {\bibfnamefont {P.}~\bibnamefont
  {Dutt}}, \bibinfo {author} {\bibfnamefont {T.~L.}\ \bibnamefont {Schmidt}},
  \bibinfo {author} {\bibfnamefont {C.}~\bibnamefont {Mora}}, \ and\ \bibinfo
  {author} {\bibfnamefont {K.}~\bibnamefont {Le~Hur}},\ }\bibfield  {title}
  {\enquote {\bibinfo {title} {{Strongly correlated dynamics in multichannel
  quantum $\ensuremath{RC}$ circuits}},}\ }\href {\doibase
  10.1103/PhysRevB.87.155134} {\bibfield  {journal} {\bibinfo  {journal} {Phys.
  Rev. B}\ }\textbf {\bibinfo {volume} {87}},\ \bibinfo {pages} {155134}
  (\bibinfo {year} {2013})}\BibitemShut {NoStop}%
\bibitem [{\citenamefont {Rodionov}\ \emph {et~al.}(2009)\citenamefont
  {Rodionov}, \citenamefont {Burmistrov},\ and\ \citenamefont
  {Ioselevich}}]{Rodionov2009}%
  \BibitemOpen
  \bibfield  {author} {\bibinfo {author} {\bibfnamefont {Ya.~I.}\ \bibnamefont
  {Rodionov}}, \bibinfo {author} {\bibfnamefont {I.~S.}\ \bibnamefont
  {Burmistrov}}, \ and\ \bibinfo {author} {\bibfnamefont {A.~S.}\ \bibnamefont
  {Ioselevich}},\ }\bibfield  {title} {\enquote {\bibinfo {title} {{Charge
  relaxation resistance in the Coulomb blockade problem}},}\ }\href {\doibase
  10.1103/PhysRevB.80.035332} {\bibfield  {journal} {\bibinfo  {journal} {Phys.
  Rev. B}\ }\textbf {\bibinfo {volume} {80}},\ \bibinfo {pages} {035332}
  (\bibinfo {year} {2009})}\BibitemShut {NoStop}%
\bibitem [{\citenamefont {Burmistrov}\ and\ \citenamefont
  {Pruisken}(2008)}]{Burmistrov2008}%
  \BibitemOpen
  \bibfield  {author} {\bibinfo {author} {\bibfnamefont {I.~S.}\ \bibnamefont
  {Burmistrov}}\ and\ \bibinfo {author} {\bibfnamefont {A.~M.~M.}\ \bibnamefont
  {Pruisken}},\ }\bibfield  {title} {\enquote {\bibinfo {title} {Coulomb
  blockade and superuniversality of the $\ensuremath{\theta}$ angle},}\ }\href
  {\doibase 10.1103/PhysRevLett.101.056801} {\bibfield  {journal} {\bibinfo
  {journal} {Phys. Rev. Lett.}\ }\textbf {\bibinfo {volume} {101}},\ \bibinfo
  {pages} {056801} (\bibinfo {year} {2008})}\BibitemShut {NoStop}%
\bibitem [{\citenamefont {Burmistrov}\ and\ \citenamefont
  {Pruisken}(2010)}]{Burmistrov2010}%
  \BibitemOpen
  \bibfield  {author} {\bibinfo {author} {\bibfnamefont {I.~S.}\ \bibnamefont
  {Burmistrov}}\ and\ \bibinfo {author} {\bibfnamefont {A.~M.~M.}\ \bibnamefont
  {Pruisken}},\ }\bibfield  {title} {\enquote {\bibinfo {title} {Macroscopic
  charge quantization in single-electron devices},}\ }\href {\doibase
  10.1103/PhysRevB.81.085428} {\bibfield  {journal} {\bibinfo  {journal} {Phys.
  Rev. B}\ }\textbf {\bibinfo {volume} {81}},\ \bibinfo {pages} {085428}
  (\bibinfo {year} {2010})}\BibitemShut {NoStop}%
\bibitem [{\citenamefont {Semenov}(2013)}]{Semenov2013}%
  \BibitemOpen
  \bibfield  {author} {\bibinfo {author} {\bibfnamefont {A.~G.}\ \bibnamefont
  {Semenov}},\ }\href@noop {} {\enquote {\bibinfo {title} {On the macroscopic
  quantization in mesoscopic rings and single-electron devices},}\ } (\bibinfo
  {year} {2013}),\ \Eprint {http://arxiv.org/abs/1307.6615} {arXiv:1307.6615}
  \BibitemShut {NoStop}%
\bibitem [{\citenamefont {Averin}\ and\ \citenamefont
  {Nazarov}(1990)}]{Averin1990}%
  \BibitemOpen
  \bibfield  {author} {\bibinfo {author} {\bibfnamefont {D.~V.}\ \bibnamefont
  {Averin}}\ and\ \bibinfo {author} {\bibfnamefont {Yu.~V.}\ \bibnamefont
  {Nazarov}},\ }\bibfield  {title} {\enquote {\bibinfo {title} {Virtual
  electron diffusion during quantum tunneling of the electric charge},}\ }\href
  {\doibase 10.1103/PhysRevLett.65.2446} {\bibfield  {journal} {\bibinfo
  {journal} {Phys. Rev. Lett.}\ }\textbf {\bibinfo {volume} {65}},\ \bibinfo
  {pages} {2446} (\bibinfo {year} {1990})}\BibitemShut {NoStop}%
\bibitem [{\citenamefont {Mezei}(1971)}]{Mezei1971}%
  \BibitemOpen
  \bibfield  {author} {\bibinfo {author} {\bibfnamefont {F.}~\bibnamefont
  {Mezei}},\ }\bibfield  {title} {\enquote {\bibinfo {title} {Theory of
  electron tunneling via real intermediate states},}\ }\href {\doibase
  10.1103/PhysRevB.4.3775} {\bibfield  {journal} {\bibinfo  {journal} {Phys.
  Rev. B}\ }\textbf {\bibinfo {volume} {4}},\ \bibinfo {pages} {3775} (\bibinfo
  {year} {1971})}\BibitemShut {NoStop}%
\bibitem [{\citenamefont {Shekhter}(1973)}]{Shekhter1972}%
  \BibitemOpen
  \bibfield  {author} {\bibinfo {author} {\bibfnamefont {R.~I.}\ \bibnamefont
  {Shekhter}},\ }\bibfield  {title} {\enquote {\bibinfo {title} {Zero anomalies
  in the resistance of a tunnel junction containing metallic inclusions in the
  oxide layer},}\ }\href@noop {} {\bibfield  {journal} {\bibinfo  {journal}
  {Sov. Phys. JETP}\ }\textbf {\bibinfo {volume} {36}},\ \bibinfo {pages} {747}
  (\bibinfo {year} {1973})}\BibitemShut {NoStop}%
\bibitem [{\citenamefont {Kulik}\ and\ \citenamefont
  {Shekhter}(1975)}]{Kulik1975}%
  \BibitemOpen
  \bibfield  {author} {\bibinfo {author} {\bibfnamefont {I.~O.}\ \bibnamefont
  {Kulik}}\ and\ \bibinfo {author} {\bibfnamefont {R.~I.}\ \bibnamefont
  {Shekhter}},\ }\bibfield  {title} {\enquote {\bibinfo {title} {Kinetic
  phenomena and charge discreteness effects in granulated media},}\ }\href@noop
  {} {\bibfield  {journal} {\bibinfo  {journal} {Sov. Phys. JETP}\ }\textbf
  {\bibinfo {volume} {41}},\ \bibinfo {pages} {308} (\bibinfo {year}
  {1975})}\BibitemShut {NoStop}%
\bibitem [{\citenamefont {Abrikosov}\ \emph {et~al.}(1963)\citenamefont
  {Abrikosov}, \citenamefont {Gorkov},\ and\ \citenamefont
  {Dzyaloshinski}}]{AGD}%
  \BibitemOpen
  \bibfield  {author} {\bibinfo {author} {\bibfnamefont {A.~A.}\ \bibnamefont
  {Abrikosov}}, \bibinfo {author} {\bibfnamefont {L.~P.}\ \bibnamefont
  {Gorkov}}, \ and\ \bibinfo {author} {\bibfnamefont {I.~E.}\ \bibnamefont
  {Dzyaloshinski}},\ }\href@noop {} {\emph {\bibinfo {title} {Methods of
  Quantum Field Theory in Statistical Physics}}}\ (\bibinfo  {publisher}
  {Prentice-Hall},\ \bibinfo {address} {Englewood Cliffs, NJ},\ \bibinfo {year}
  {1963})\BibitemShut {NoStop}%
\bibitem [{\citenamefont {Abrikosov}(1965)}]{abrikosov}%
  \BibitemOpen
  \bibfield  {author} {\bibinfo {author} {\bibfnamefont {A.~A.}\ \bibnamefont
  {Abrikosov}},\ }\bibfield  {title} {\enquote {\bibinfo {title} {Electron
  scattering on magnetic impurities in metals and anomalous resistivity
  effects},}\ }\href@noop {} {\bibfield  {journal} {\bibinfo  {journal}
  {Physics. Physique. Fizika.}\ }\textbf {\bibinfo {volume} {2}},\ \bibinfo
  {pages} {5} (\bibinfo {year} {1965})}\BibitemShut {NoStop}%
\bibitem [{\citenamefont {Larkin}\ and\ \citenamefont
  {Melnikov}(1972)}]{larkin}%
  \BibitemOpen
  \bibfield  {author} {\bibinfo {author} {\bibfnamefont {A.~I.}\ \bibnamefont
  {Larkin}}\ and\ \bibinfo {author} {\bibfnamefont {V.~I.}\ \bibnamefont
  {Melnikov}},\ }\bibfield  {title} {\enquote {\bibinfo {title} {Magnetic
  impurities in an almost magnetic metal},}\ }\href@noop {} {\bibfield
  {journal} {\bibinfo  {journal} {Sov. Phys. JETP}\ }\textbf {\bibinfo {volume}
  {34}},\ \bibinfo {pages} {656} (\bibinfo {year} {1972})}\BibitemShut
  {NoStop}%
\bibitem [{\citenamefont {Sachdev}\ and\ \citenamefont {Ye}(1993)}]{Sachdev}%
  \BibitemOpen
  \bibfield  {author} {\bibinfo {author} {\bibfnamefont {S.}~\bibnamefont
  {Sachdev}}\ and\ \bibinfo {author} {\bibfnamefont {J.}~\bibnamefont {Ye}},\
  }\bibfield  {title} {\enquote {\bibinfo {title} {Gapless spin-fluid ground
  state in a random quantum {Heisenberg} magnet},}\ }\href {\doibase
  10.1103/PhysRevLett.70.3339} {\bibfield  {journal} {\bibinfo  {journal}
  {Phys. Rev. Lett.}\ }\textbf {\bibinfo {volume} {70}},\ \bibinfo {pages}
  {3339} (\bibinfo {year} {1993})}\BibitemShut {NoStop}%
\bibitem [{\citenamefont {Zhu}\ and\ \citenamefont {Si}(2002)}]{Si}%
  \BibitemOpen
  \bibfield  {author} {\bibinfo {author} {\bibfnamefont {L.}~\bibnamefont
  {Zhu}}\ and\ \bibinfo {author} {\bibfnamefont {Q.}~\bibnamefont {Si}},\
  }\bibfield  {title} {\enquote {\bibinfo {title} {Critical local-moment
  fluctuations in the {Bose-Fermi Kondo} model},}\ }\href {\doibase
  10.1103/PhysRevB.66.024426} {\bibfield  {journal} {\bibinfo  {journal} {Phys.
  Rev. B}\ }\textbf {\bibinfo {volume} {66}},\ \bibinfo {pages} {024426}
  (\bibinfo {year} {2002})}\BibitemShut {NoStop}%
\bibitem [{\citenamefont {Zar\'and}\ and\ \citenamefont
  {Demler}(2002)}]{Demler}%
  \BibitemOpen
  \bibfield  {author} {\bibinfo {author} {\bibfnamefont {G.}~\bibnamefont
  {Zar\'and}}\ and\ \bibinfo {author} {\bibfnamefont {E.}~\bibnamefont
  {Demler}},\ }\bibfield  {title} {\enquote {\bibinfo {title} {Quantum phase
  transitions in the {Bose-Fermi Kondo} model},}\ }\href {\doibase
  10.1103/PhysRevB.66.024427} {\bibfield  {journal} {\bibinfo  {journal} {Phys.
  Rev. B}\ }\textbf {\bibinfo {volume} {66}},\ \bibinfo {pages} {024427}
  (\bibinfo {year} {2002})}\BibitemShut {NoStop}%
\bibitem [{\citenamefont {Schoeller}\ and\ \citenamefont
  {Sch\"on}(1994)}]{schoeller}%
  \BibitemOpen
  \bibfield  {author} {\bibinfo {author} {\bibfnamefont {H.}~\bibnamefont
  {Schoeller}}\ and\ \bibinfo {author} {\bibfnamefont {G.}~\bibnamefont
  {Sch\"on}},\ }\bibfield  {title} {\enquote {\bibinfo {title} {{Mesoscopic
  quantum transport: Resonant tunneling in the presence of a strong Coulomb
  interaction}},}\ }\href {\doibase 10.1103/PhysRevB.50.18436} {\bibfield
  {journal} {\bibinfo  {journal} {Phys. Rev. B}\ }\textbf {\bibinfo {volume}
  {50}},\ \bibinfo {pages} {18436} (\bibinfo {year} {1994})}\BibitemShut
  {NoStop}%
\bibitem [{\citenamefont {Zar\'and}\ \emph {et~al.}(2000)\citenamefont
  {Zar\'and}, \citenamefont {Zim\'anyi},\ and\ \citenamefont
  {Wilhelm}}]{zarand}%
  \BibitemOpen
  \bibfield  {author} {\bibinfo {author} {\bibfnamefont {G.}~\bibnamefont
  {Zar\'and}}, \bibinfo {author} {\bibfnamefont {G.~T.}\ \bibnamefont
  {Zim\'anyi}}, \ and\ \bibinfo {author} {\bibfnamefont {F.}~\bibnamefont
  {Wilhelm}},\ }\bibfield  {title} {\enquote {\bibinfo {title} {{Two-channel
  versus infinite-channel Kondo models for the single-electron transistor}},}\
  }\href {\doibase 10.1103/PhysRevB.62.8137} {\bibfield  {journal} {\bibinfo
  {journal} {Phys. Rev. B}\ }\textbf {\bibinfo {volume} {62}},\ \bibinfo
  {pages} {8137} (\bibinfo {year} {2000})}\BibitemShut {NoStop}%
\bibitem [{\citenamefont {Rodionov}\ \emph {et~al.}(2010)\citenamefont
  {Rodionov}, \citenamefont {Burmistrov},\ and\ \citenamefont
  {Chtchelkatchev}}]{Rodionov2010a}%
  \BibitemOpen
  \bibfield  {author} {\bibinfo {author} {\bibfnamefont {Ya.~I.}\ \bibnamefont
  {Rodionov}}, \bibinfo {author} {\bibfnamefont {I.~S.}\ \bibnamefont
  {Burmistrov}}, \ and\ \bibinfo {author} {\bibfnamefont {N.~M.}\ \bibnamefont
  {Chtchelkatchev}},\ }\bibfield  {title} {\enquote {\bibinfo {title}
  {{Relaxation dynamics of the electron distribution in the Coulomb-blockade
  problem}},}\ }\href {\doibase 10.1103/PhysRevB.82.155317} {\bibfield
  {journal} {\bibinfo  {journal} {Phys. Rev. B}\ }\textbf {\bibinfo {volume}
  {82}},\ \bibinfo {pages} {155317} (\bibinfo {year} {2010})}\BibitemShut
  {NoStop}%
\bibitem [{\citenamefont {Rodionov}\ and\ \citenamefont
  {Burmistrov}(2010)}]{Rodionov2010b}%
  \BibitemOpen
  \bibfield  {author} {\bibinfo {author} {\bibfnamefont {Ya.~I.}\ \bibnamefont
  {Rodionov}}\ and\ \bibinfo {author} {\bibfnamefont {I.~S.}\ \bibnamefont
  {Burmistrov}},\ }\bibfield  {title} {\enquote {\bibinfo {title}
  {Out-of-equilibrium admittance of single electron box under strong {Coulomb}
  blockade},}\ }\href@noop {} {\bibfield  {journal} {\bibinfo  {journal} {JETP
  Lett.}\ }\textbf {\bibinfo {volume} {92}},\ \bibinfo {pages} {696} (\bibinfo
  {year} {2010})}\BibitemShut {NoStop}%
\bibitem [{\citenamefont {Rodionov}\ and\ \citenamefont
  {Burmistrov}()}]{Future}%
  \BibitemOpen
  \bibfield  {author} {\bibinfo {author} {\bibfnamefont {Ya.~I.}\ \bibnamefont
  {Rodionov}}\ and\ \bibinfo {author} {\bibfnamefont {I.~S.}\ \bibnamefont
  {Burmistrov}},\ }\href@noop {} {}\bibinfo {note} {{unpublished}}\BibitemShut
  {NoStop}%
\bibitem [{\citenamefont {Ben-Jacob}\ \emph {et~al.}(1983)\citenamefont
  {Ben-Jacob}, \citenamefont {Mottola},\ and\ \citenamefont
  {Sch\"on}}]{Ben-Jacob1983}%
  \BibitemOpen
  \bibfield  {author} {\bibinfo {author} {\bibfnamefont {Eshel}\ \bibnamefont
  {Ben-Jacob}}, \bibinfo {author} {\bibfnamefont {Emil}\ \bibnamefont
  {Mottola}}, \ and\ \bibinfo {author} {\bibfnamefont {Gerd}\ \bibnamefont
  {Sch\"on}},\ }\bibfield  {title} {\enquote {\bibinfo {title} {Quantum shot
  noise in tunnel junctions},}\ }\href@noop {} {\bibfield  {journal} {\bibinfo
  {journal} {Phys. Rev. Lett.}\ }\textbf {\bibinfo {volume} {51}},\ \bibinfo
  {pages} {2064} (\bibinfo {year} {1983})}\BibitemShut {NoStop}%
\end{thebibliography}%

\end{document}